\documentclass[journal,transmag]{IEEEtran}
\usepackage{caption}
\usepackage[cmex10]{amsmath}
\usepackage{amsfonts}
\usepackage{algorithm, algorithmic}
\usepackage{array}
\usepackage{etoolbox,xstring,mfirstuc,textcase}
\usepackage{mdwmath}
\usepackage{mdwtab}
\usepackage{eqparbox}
\usepackage{mathrsfs}
\usepackage{tabularx}
\usepackage{amsthm}
\usepackage{amsmath,bm}
\usepackage{booktabs}
\usepackage{multirow}
\usepackage{diagbox}
\usepackage{txfonts}
\usepackage{bm}
\usepackage{makecell}
\usepackage{subfigure}
\usepackage{graphicx}
\usepackage{enumitem}
\usepackage{cite}
\usepackage{ltxtable}
\usepackage{amsmath}
\usepackage{amssymb}
\usepackage{mathrsfs}
\usepackage{setspace}
\usepackage{multirow}
\usepackage{subfigure}

\usepackage{acronym}
\usepackage{easyReview}
\usepackage[normalem]{ulem} 

\usepackage[utf8]{inputenc}
\usepackage[english]{babel}
\def\R{\mathbb{R}}

\usepackage{dirtytalk}
\usepackage{multicol}
\usepackage[none]{hyphenat}

\title{A Gaussian Process Regression based  Dynamical Models Learning Algorithm for Target Tracking}

\author{Mengwei Sun, \IEEEmembership{Member, IEEE}, 
Mike E. Davies, \IEEEmembership{Fellow, IEEE}, Ian K. Proudler, James R. Hopgood, \IEEEmembership{Member, IEEE}% and Third C. Author, Jr., \IEEEmembership{Member, IEEE}
\thanks{M. W. Sun, M. E. Davies and J. R. Hopgood are with Institute 
of Digital Communications, University of Edinburgh, Edinburgh, EH9 3FG, U.K. E-mail: (msun; mike.davies; james.hopgood)@ed.ac.uk.

I. Proudler is with  the Centre for Signal \& Image Processing (CeSIP), Department of Electronic \& Electrical Engineering, University of Strathclyde, Glasgow, G1 1XW, U.K. E-mail: 
ian.proudler@strath.ac.uk.

This work was supported by the Engineering and Physical Sciences Research Council (EPSRC) Grant number EP/S000631/1; and the MOD University Defence Research Collaboration (UDRC) in Signal Processing.
}
}

\begin{document}

\maketitle
%\tableofcontents

\acrodef{MM}{multiple model}
\acrodef{GP}{Gaussian process}
\acrodef{GPR}{Gaussian process regression}
\acrodef{PF} {particle filter}
\acrodef{JPDA} {joint probabilistic data association} 
\acrodef{MP} {message passing}
\acrodef{MTT} {multi-target tracking}
\acrodef{STT} {single-target tracking}
\acrodef{PDF} {probability density function}
\acrodef{RFS}{random finite set} 
\acrodef{CV}{constant velocity} 
\acrodef{CA}{constant acceleration} 
\acrodef{CT}{coordinated turn}
\acrodef{CNN}{convolutional neural network}
\acrodef{DA}{data association}
\acrodef{LSTM}{long short-term memory}
\acrodef{JPDAF}{joint probabilistic data association filter}
\acrodef{MHT}{multi-hypothesis tracker }
\acrodef{MAP}{maximum a posteriori }
\acrodef{SPA}{sum–product algorithm }
\acrodef{DSSM}{dynamic state-space model}
\acrodef{AWGN}{additive white Gaussian noise}
\acrodef{SE}{squared exponential}
\acrodef{LR}{likelihood ratio}
\acrodef{RMSE}{root mean square error }
\acrodef{IMM-PF}{interacting multiple model-particle filter }
\acrodef{IMM}{interacting multiple model}
\acrodef{GOSAP}{generalized optimal sub-patten assignment metric}
\acrodef{DA}{data association}
\acrodef{MAP}{maximum a posteriori}
\acrodef{BP}{belief propagation}
\acrodef{EKF}{extended Kalman filter}
\acrodef{UKF}{unscented Kalman filter}
\acrodef{KF}{Kalman filter}
\acrodef{GCT}{Gradual coordinated turns}
\acrodef{pdf}{probability density function}
\acrodef{MC}{Monte Carlo }
\acrodef{NSIM}{naturally shift invariant motion}

\begin{abstract}

Maneuvering target tracking is a challenging problem for sensor systems because of the unpredictability of the targets’ motions. {This paper proposes a novel data-driven method for learning the dynamical motion model of a target. Non-parametric \acf{GPR} is used to learn a target's  \ac{NSIM} behavior, which is translationally invariant and does not need to be constantly updated as the target moves. 
The learned Gaussian processes (GPs) can be applied to track targets within different surveillance regions from the surveillance region of the training data by being incorporated into the \ac{PF} implementation.} The performance of our proposed approach is evaluated over different maneuvering scenarios by being compared with commonly used \ac{IMM}--\ac{PF} methods and  provides around 90\% performance improvement 
for a \ac{MTT}  highly maneuvering scenario.
\end{abstract} 

\acresetall

\section{Introduction}

\IEEEPARstart{M}{aneuvering} target 
tracking is a fundamental task in sensor-based applications, such as radar, sonar, and navigation \cite{LIU2016183, 6507656, 1413764}. 
Based on the Bayesian framework, target tracking is usually solved via a recursive update of the posterior \acf{pdf} of the target states. 
The application of Bayesian filters is based on dynamical motion models and  sensor measurement models \cite{1413764, 5730505}. 
However, there may be significant motion model uncertainty when targets undergo unknown or mixed maneuvering behaviors \cite{1561886, JLYang}, such that the evolution of the target state is too complex to be approximated by a reasonable number of mathematical models. The uncertainty in motion behavior can also be due to various parameters in generative models not being known {\textit{a priori}} or if they are time-varying. The tracking performance of Bayesian filters would degrade and even become unacceptable if incorrect models or parameters are applied. This paper focuses on the issue of learning uncertain motion models with higher flexibility and incorporating the learned models into the Bayesian framework for target tracking.

\subsection{State-of-the-Art in Bayesian Filters} 

\Ac{MM} methods are commonly used  to deal with the model uncertainty problem \cite{1561886}. The basic idea of \ac{MM} methods is to use a bank of elemental filters with different motion models to capture the mixed motion behaviors. Then, the overall estimate is generated based on the results achieved by each elemental filter \cite{1561886}. 
For the cases when the target's behavior is a time-varying or doubly-stochastic process, 
\cite{JLYang} and \cite{LiuZX} incorporated adaptive parameter estimation methods into the Bayesian filter framework to deal with the uncertainty of unknown parameters, where the unknown parameter is estimated by approximating its distribution with particles and corresponding weights. \Ac{MM} methods  may be limited in performance if the assumed trajectories are not modeled accurately enough to capture all aspects of the motion behaviors, and the estimation of unknown parameters results in increased computation complexity.

\subsection{State-of-the-Art in Machine Learning Approaches} 
As an alternative to traditional Bayesian methods, machine learning-based tracking algorithms have been proposed recently \cite{8099886, 7088657, 8096087, 9011258,Batch_nonlinear, 9185002,9272174}. In \cite{8099886}, a quadruplet \acf{CNN} based algorithm was designed to learn object association for multi-object tracking via 2D image processing.
An extended target tracking method using a \acf{GP} was proposed in \cite{7088657} and \cite{8096087} to track the irregularly shaped object whose moving trajectory follows a fixed motion model. These papers use a \ac{GP} for shape modeling but not for capturing the motion behavior. A \acf{LSTM} neural network was used in \cite{9011258} to perform the prediction step of target localization. The proposed filter with \ac{LSTM} prediction can improve the tracking accuracy by using computationally efficient low-dimensional state spaces but only considers the \ac{STT} scenario. In \cite{9011258}, the continuous-discrete estimation of the target’s trajectory is viewed as a one-dimensional sparse GP, with time as the independent variable and measurements acquired at discrete times. Recursive GP-based approaches for target tracking and smoothing were developed in \cite{Batch_nonlinear, 9185002}, with online training and parameter learning. Only \ac{STT} with a linear observation model (i.e. $X$-axis and $Y$-axis coordinates)
{is} considered. Furthermore, it was assumed that the motion model could be decomposed into separate models for the X and Y axes. In \cite{9272174}, the motion behavior of a single target is modeled as a Markov process which switches between the \ac{CV} and \ac{CT} models with the specified transition probabilities. 
The complex mixed model is learned as joint \acp{GP} for position and velocity prediction. Then the trained \acp{GP} are integrated into the PF for tracking the maneuvering targets in real-time. 

The learning-based tracking algorithms proposed in \cite{Batch_nonlinear, 9185002, 9272174} were for motion models based on Cartesian positions which
are problematic because positions are not translationally invariant,
i.e. the motion model needs to be constantly updated as the
target moves. {This paper proposes learning a \acf{NSIM} model. Specifically, the change in
the target’s Cartesian positions on the $X$-axis and $Y$-axis in one
instant (i.e. Cartesian velocities), instead of the Cartesian
positions, is learned as multi-independent \acp{GP}.}

\subsection{ Proposed Method and Contributions}
{
This work follows
 a learning-to-track framework for tracking problems with uncertain motion models and proposes a new learning method with which the motion behaviors 
of targets are learned as \acp{GP} based on a training data set. Then we show how the learned motion models are integrated into the \ac{PF} and \ac{BP} to realize target tracking. The  three main  contributions  are:
\begin{itemize}
    \item  We propose a new \ac{GP}-based dynamical motion model learning method. The \ac{NSIM} behavior of a target is learned using \acf{GPR} based on the Cartesian velocities.
   The learned  \acp{GP}  are translationally invariant and can be applied to track objects within different surveillance regions from the surveillance region of the training data.
    \item We explore the potential of using the learned motion model within the \ac{PF}  \cite{978374, article} for \ac{STT} and \ac{PF} based \ac{BP} \cite{910572, koller2009probabilistic, 6301723} for \acf{MTT} to demonstrate that the proposed algorithm  is a plug-and-play learning method in target tracking systems.
    \item The proposed learning method is firstly tested in \ac{STT} with different motion scenarios to demonstrate its capability to 
    learn uncertain motion models with higher flexibility. Then, the learned motion model is plugged into \ac{MTT}, which shows significant improvement in tracking performance compared with \acf{IMM}-\ac{PF} methods.  Additionally, offline learning is performed only once and then applied for \ac{MTT}. Hence, the online tracking time is not affected by the learning process.
\end{itemize}

Compared with the available learning-based tracking methods using target positions, the motivations and advantages of
the proposed \ac{GP}-based \ac{NSIM} model include:
\begin{itemize}
\item The information available for use in the tracking problems may be limited depending on how we envisage getting the training data. For example, if the data was taken from other previously tracked targets, it could be challenging to access speed or acceleration.
 \item  The existing methods learn the motion models based on target positions, which are not translationally invariant and need to update the model as the target moves constantly. Moreover, the tested targets' surveillance regions may differ from the training data's surveillance region. On the other hand, the motion of an object is usually controlled by the application of a force, e.g. aircraft control surfaces. As such, the motion is translationally invariant. Therefore, the proposed \ac{NSIM} model based on Cartesian velocities will be translationally invariant and can reduce the requirement of training data size and algorithm complexity.
 \item The learned GP-based NSIM model can provide confidence intervals for particle sampling in the prediction step of the \ac{PF} to provide a reliable estimate of model uncertainty.
\end{itemize}}

 \subsection{Paper Organization and Notations}

The remainder of this paper is organized as follows. The problem solved in this investigation is defined in Section II. Section III reviews the \ac{GPR} and proposes the GP-based learning method of \ac{NSIM} behavior. The proposed algorithm, which incorporates the learned \acp{GP} into the target tracking, is developed in Section IV. Simulation results and
performance comparisons are presented in Section V. Conclusions are provided in Section VI. This paper presents matrices with uppercase letters, vectors with boldface lowercase letters and scalars with lowercase letters. The notations are defined and summarized in Table I.
\begin{table}[!h]
{	\small
	 \caption{The meaning of notations.}
\label{tab:test}
\begin{center}
\begin{tabular}{|c|c|}
	\hline
\textbf{Notation} & \textbf{Meaning} \\
\hline
$\mathbf{x}^*_{1:N}$
& Target states of training data sequence
\\
\hline
$\mathbf{y}^*_{1:N}$
& Observations of training data sequence
\\
\hline
$\mathbf{x}_{1:T}$
& Target states of {test} data sequence
\\
\hline
$\mathbf{y}_{1:T}$
& Observations of {test}  data sequence
\\
\hline
$n = 0,\dots, N-1$
& Training time index
\\
\hline
$t = 1,\dots, T$
& Test tracking time index
\\
\hline
$k = 1,\dots, K$
& Target index
\\
\hline
$\left[\xi_{t,k},  \eta_{t,k}\right]$
& Cartesian positions  at 2-D plane
\\
\hline
$\left[\Delta \xi_{t,k},  \Delta \eta_{t,k}\right]$
& Cartesian velocities at 2-D plane
\\
\hline
$v_{t,k}$
& Ground speed
\\
\hline
$\phi_{t,k}$
& Velocity heading angle
\\
\hline
${\alpha}^t_{t,k}$
& Along-track acceleration 
\\
\hline
${\alpha}^n_{t,k}$
& Cross-track acceleration
\\
\hline
$Q$
& Covariance matrix of process noise
\\
\hline
$R$
& Covariance matrix of measurement noise
\\
\hline
$\mathcal{D} $
& Training data set 
\\
\hline
${f}(\cdot)$
& Motion function
\\
\hline$h(\cdot)$
& Measurement function
\\
\hline
${\dot{f}}^{*\xi}(\cdot)$ 
&$X$-axis velocity motion function  for training
\\
\hline
${\dot{f}}^{*\eta}(\cdot)$ 
&$Y$-axis velocity motion function  for training 
\\
\hline
${\dot{f}}^{\xi}(\cdot)$ 
&$X$-axis velocity  motion function  for testing
\\
\hline
${\dot{f}}^{\eta}(\cdot)$ 
&$Y$-axis velocity motion function  for testing
\\
\hline
\end{tabular}
\end{center}
}
\end{table}

\section{Problem Definition and Background}

In this section, we will introduce the system model  and  conventional Bayesian approaches of target tracking.
 
\subsection{System Model}
The \acf{DSSM} for target tracking is given by the following motion and measurement models:
\begin{align}
\label{eq:MTT-MOTION}
    \mathbf{x}_{t, k}  &= f_{k}(\mathbf{x}_{t-1,k}, \mathbf{v}_{t-1,k})\\
  \label{eq:MTT-MEA}  
    \mathbf{y}_{t,k} &= h_{k}( \mathbf{x}_{t,k}, \mathbf{e}_{t,k}).
\end{align}
Here, $t$ represents the time index, $t = 1 , \dots, T$,  $k$ is the target index, $k = 1,\dots, K$. %where $K$ is the number of targets which  is set to be fixed and known \textit{a priori} in this investigation. 
For \ac{STT} scenarios, $K = 1$.
The state motion model and the measurement model are represented as  $f_k: \R^{n_x} \times \R^{n_v} \rightarrow \R^{n_x}$ and {$h_{k}:  \R^{n_x} \times \R^{n_e} \rightarrow \R^{n_y}$,} respectively,  where $n_x$, $n_v$, $n_y$ and $n_e$  are dimensions of the state, process noise, observation, and measurement noise vectors, respectively. The process noise $ \mathbf{v}_{t,k}$ and the measurement noise $\mathbf{e}_{t,k}$ are assumed to be \acf{AWGN}.
At each time instant, the sensor node detects each target with detection probability $P_D \leq 1$. Owing to missed detections and clutter, the number of detected measurements at time $t$ may not be equal to  $K$. $\mathcal{K}_t$ is defined as the number of measurements at time $t$ and is time-varying.  The set of measurements detected at time $t$ is denoted as ${Y}_t = \{\mathbf{y}_{t,1}, \dots, \mathbf{y}_{t,\mathcal{K}_t}\}$. The  measurement index is represented as $\kappa$ and $\kappa = 1,\dots, \mathcal{K}_t$.
The number of clutter points is modeled as a Poisson point process with intensity  $\lambda_{\mathrm{fa}}$. The corresponding clutter measurements are independent and identically distributed (iid) with \ac{pdf} $p_c(\mathbf{y}_{t,\kappa})$. 

%  and denoted by  $\mathbf{v}_{t,k} \sim \mathcal{N} ({\mathbf{0}}, {Q})$ and $\mathbf{e}_{t,k}\sim \mathcal{N} ({\mathbf{0}}, {R})$, where $Q$ and $R$ represent the covariance of process and measurement noise, respectively,

%Validated measurements are defined as those detected measurements whose predictive likelihood are larger than a certain  gate  probability $p_g$ \cite{6302287}, i.e.
%\begin{align}
%\label{validation_region}
%   Y_t = \left\lbrace \mathbf{y}_{t,\kappa}:
%\mathcal{L}_k(\kappa) = p(\mathbf{y}_{t,\kappa}\mid %\hat{\mathbf{x}}_{t-1,k}) \geq p_g \right\rbrace.
%\end{align}

%two necessary steps are processed before state estimations, i.e.  measurement validation and data association. The measurement validation aims to determine the validated measurement set which are associated with the targets.

\subsection{Sequential Bayesian Filters for Target Tracking}

Based on the Bayesian theorem, most target tracking methods are sequential Bayesian filters which consist of the prediction and the update steps.
 In the prediction step, the  prior distribution of targets' states at time $t$ is calculated via the Chapman–Kolmogorov equation as \cite{8290605}
\begin{align}
\label{eq:eqBayes_prior}
\begin{split}
 &p(\mathbf{x}_{t,1:K}\mid {Y_{1:t-1}})\\
 &\qquad=\int {p(\mathbf{x}_{t,1:K} \mid \mathbf{x}_{t-1,1:K})p(\mathbf{x}_{t-1,1:K}\mid {Y_{1:t-1}}) } d\mathbf{x}_{t-1,1:K}.
    \end{split}
\end{align}
In the update step, the joint posterior \ac{pdf} of the  targets' states is calculated  as \cite{8290605}
\begin{align}
\label{eq:eqBayes}
    p(\mathbf{x}_{t,1:K}\mid {Y_{1:t}})  
    = \frac{p(Y_t\mid \mathbf{x}_{t,1:K})p(\mathbf{x}_{t,1:K}\mid {Y_{1:t-1}}) }{\int p(Y_t\mid \mathbf{x}_{t,1:K})p(\mathbf{x}_{t,1:K}\mid {Y_{1:t-1}}) d \mathbf{x}_{t,1:K}}.
\end{align}
%For the \ac{STT}, the estimation in \eqref{eq:eqBayes} can be realized in a sequential way as
%\begin{align}
%\label{eq:eqBayesSTT}
%    p(\mathbf{x}_{t}|{\mathbf{y}_{1:t}})  
%    = \frac{p(\mathbf{y}_t\mid \mathbf{x}_{t}) p(\mathbf{x}_{t}\mid {\mathbf{y}_{1:t-1}}) }{\int p(\mathbf{y}_t\mid \mathbf{x}_{t}) p(\mathbf{x}_{t}\mid  {\mathbf{y}_{1:t-1}})d \mathbf{x}_{t}}.
%\end{align}

 For \ac{STT} in non-linear systems, the \ac{PF} uses random particles with associated weights to approximate the posterior \ac{pdf}
as 
\begin{align}
    p(\mathbf{x}_{t}\mid {Y_{1:t}}) \approx \sum_{m=1}^M w_n^{\{m\}} \delta\left(\mathbf{x}_t - \mathbf{x}_t^{\{m\}}\right)
\end{align}
where $\delta(\cdot)$ is the Dirac delta function, the number of particles used at a given time is represented as $M$. The particles $\mathbf{x}_t^{\{1:M\}}$ are drawn from an importance density which can be set as  $\mathbf{x}_t^{\{m\}} = f(\mathbf{x}_{t-1}^{\{m\}}, \mathbf{v}_{t-1}^{\{m\}})$\cite{978374}.

\begin{figure}[!t]
	\centering
	\includegraphics [width=0.45\textwidth]{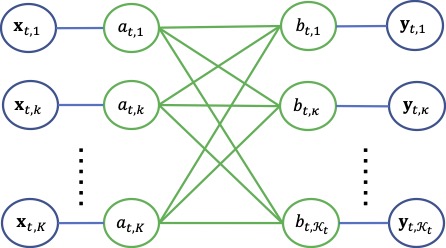}
	\caption{{\footnotesize{Factor graphs for single-sensor MTT, $\mathbf{a}_{t} = \left[{a}_{t,1},\dots,{a}_{t,K}\right]^{\rm{T}}$  is the target-oriented association vector and  $\mathbf{b}_{t} = \left[{b}_{t,1},\dots,{b}_{t,\mathcal{K}_t}\right]^{\rm{T}}$ is the measurement-oriented association vector. }}}
	\label{fig1}
\end{figure}

For \ac{MTT}, the association between targets and measurements is described by two \ac{DA} vectors shown in Fig. \ref{fig1} \cite{6978890,7889057}, i.e.  the target-oriented association vector
$\mathbf{a}_{t} = \left[{a}_{t,1},\dots,{a}_{t,K}\right]^{\rm{T}}$ and the measurement-oriented association vector $\mathbf{b}_{t} = \left[ {b}_{t,1},\dots,{b}_{t,\mathcal{K}_t} \right]^{\rm{T}}$ where $\mathcal{K}_t$ is the number of measurements at time $t$.  The \ac{DA} vectors are defined as \cite{6302287}
\begin{itemize}
    \item ${a}_{t,k} \triangleq \kappa \in \{1,\dots,\mathcal{K}_t\} $ if at time $t$, the  $k$-th target  generates the $\kappa$-th measurement.
    \item  ${a}_{t,k}\triangleq \kappa  = 0 $ if  at time $t$, the $k$-th target does not generate a measurement, i.e. {missed detection} occurs.
    \item ${b}_{t,\kappa} \triangleq k \in \{1,\dots,K\} $ if at time $t$, the $\kappa$-th measurement is hypothesized to be associated with $k$-th target.
    \item  ${b}_{t,\kappa} \triangleq k = 0 $ if  at time $t$, the $\kappa$-th measurement is {hypothesized not to be} generated by a target, i.e. clutter.
\end{itemize}
The objective of MTT is to calculate the marginal  posterior \ac{pdf} $p(\mathbf{x}_{t,k}\mid Y_{1:t}), \, k= 1,\dots,K$ by
\begin{equation}
\label{eq:eq_marginal_pdf}
   \begin{aligned}
   p(\mathbf{x}_{t,k}\mid Y_{1:t})=\sum_{a_{t,k}=0}^{\mathcal{K}_t}p(\mathbf{x}_{t,k}\mid a_{t,k}, Y_{1:t})p(a_{t,k}\mid Y_{1:t})
   \end{aligned} 
\end{equation}
where the marginal association probability, i.e. $p(a_{t,k}\mid Y_{1:t})$,  approximates based on the loopy \ac{BP}  algorithm, as {given by} Eqn. (22)  in \cite{6978890}.

It should be noted that most of the existing \ac{STT}/\ac{MTT} filters are fully model-driven and, therefore, of use when the  \ac{DSSM} is explicitly available. However, assuming that the state motion function is given as a simple deterministic model is restrictive. These filters are intractable for some applications where the motion model is very complicated. Instead, information about the state transitions can be inferred from examples of preceding states and current states pairs. Hence, we assume that the motion model is learned using the non-parametric \acf{GPR} \cite{books/lib/RasmussenW06}. The measurement model is deterministic by different sensor types with no correlation over time and is in a known parametric form, i.e. we know the acquisition model. In the next section, we will discuss the application of \ac{GPR} to learn motion behavior.

\section{Gaussian Progress Regression for Motion Behavior Learning}

GP models are widely used to perform Bayesian non-linear regression. A \ac{GP} model takes a training data set and a kernel as the input and outputs a joint Gaussian distribution that characterizes the value of the function at some given points \cite{phdthesisgp}. {One practical advantage of \ac{GP} is having confidence intervals for predictions that provide a reliable estimate of uncertainty.} In this section, we briefly review the standard theory of GPR and then introduce the proposed GP-based method for motion behavior learning.

\subsection{Gaussian Process Regression}

The \ac{GP} model for regression can be considered as a distribution over a function $g(\mathbf{x})$
and is defined as \cite{7088657}
\begin{align}
{g}(\mathbf{x}) & \sim \mathcal{GP}(\mathbf{m}({\mathbf{x}}), \mathit{k}({\mathbf{x}}, {\mathbf{x}}') )
\end{align}
where $\mathbf{x}$ and ${\mathbf{x}}'$ represent the input variables. 
The mean function $\mathbf{m}({\mathbf{x}})$ and the covariance function $k({{\mathbf{x}}},{\mathbf{x}}')$ of two variables $\mathbf{x}$ and ${\mathbf{x}}'$ are defined as  \cite{8882261} 
\begin{align}
\mathbf{m}({\mathbf{x}}) &= \mathbb{E}\left[{g}({\mathbf{x}})  \right]\\
  k({{\mathbf{x}}}, {\mathbf{x}}')& =  \mathbb{E}\left[\left( {g}({\mathbf{x}})- {\mathbf{m}}({\mathbf{x}}) \right)\left({g}({\mathbf{x}}')- \mathbf{m}({\mathbf{x}}') \right)\right].
\end{align}

Let us consider a general \ac{GPR}  problem with noisy observations from an unknown function described as
\begin{equation}
    z=g(\mathbf{x})+\mathbf{v}, \qquad \mathbf{v} \sim \mathcal{N}(0,\sigma^2_v {I}).
\end{equation}
{Here, $\sigma^2_v $ is the variance of the noise.} The training data set is denoted as $\mathcal{D} = \{{X}^*, \mathbf{z}^*\}$,
where ${X}^* = \left[{\mathbf{x}}_0^*, {\mathbf{x}}_1^*, \dots, {\mathbf{x}}_{N-1}^* \right]^{\rm{T}}$ are the inputs with the corresponding outputs  $\mathbf{z}^* = \left[z_0^*, {z}_1^*, \dots, {z}_{N-1}^* \right]^{\rm{T}}$. The purpose of \ac{GPR} is to derive the latent distributions of the {vector} ${\mathbf{g}} = \left[ g(\mathbf{x}_1), g(\mathbf{x}_2), \dots, g(\mathbf{x}_{T})\right]^{\rm{T}}$ at the test inputs $X = \left[ {\mathbf{x}}_1, {\mathbf{x}}_2,\dots,{\mathbf{x}}_{T}\right]^{\rm{T}}$, conditioned on the 
test measurements $\mathbf{z} = \left[ {z}_1, {z}_2,\dots ,z_{T}\right]^{\rm{T}}$ and 
{the training data} set $\mathcal{D}$. 
Specifically, the joint distribution of the training measurements $\mathbf{z}^*$ and the function at one test point, i.e. ${g}_t \triangleq g({\mathbf{x}}_{t})$, {is assumed to be Gaussian and is given by}
\begin{align}
    \begin{bmatrix}
    \mathbf{z}^*\\
    g_t
    \end{bmatrix}
    \sim \mathcal{N} \left(
    \begin{bmatrix}
    \mathbf{m}(X^*)\\
    {m}({\mathbf{x}}_t)
    \end{bmatrix},
        \begin{bmatrix}
   K(X^*,X^*)+\sigma_v^2I  &\mathbf{k}_t(X^*,{\mathbf{x}}_t)\\
   \mathbf{k}_t({\mathbf{x}}_t,X^*) &k({\mathbf{x}}_t,{\mathbf{x}}_t) 
    \end{bmatrix}
    \right).
    \label{eq:eqGP1}
\end{align}
Here, ${\mathbf{k}}_t(X^*,{\mathbf{x}}_t) = [k({\mathbf{x}}_0^*,{\mathbf{x}}_t), \dots,k({\mathbf{x}}^*_{N-1},{\mathbf{x}}_t)]^{\rm{T}}$, $k({\mathbf{x}}_t,{\mathbf{x}}_t)$ is the covariance of ${\mathbf{x}}_t$, and $K(X^*,X^*)$ denotes the covariance matrix for the training input data \cite{9272174}. The conditional distribution of the function $g(\mathbf{x}_t)$ is then derived from the joint density in \eqref{eq:eqGP1} and written 
as the following standard forms:
%\begin{align}
%K(X,X) = \begin{bmatrix}
%k({\mathbf{x}}_0,{\mathbf{x}}_0), &\dots, &k({\mathbf{x}}_0,{\mathbf{x}}_{N-1})\\
%\vdots   & \ddots & \vdots\\
%k({\mathbf{x}}_{N-1},{\mathbf{x}}_0), &\dots, &k({\mathbf{x}}_{N-1},{\mathbf{x}}_{N-1})
%\end{bmatrix}.
%\end{align}
\begin{subequations}
\begin{gather}
    ({g}_t|\mathcal{D})\sim 
\mathcal{N}(\mathbf{m}({g}_t), \mathbb{V}\left[{g}_t \right])\\
    \mathbf{m}({{g}}_t) = {m}({\mathbf{x}}_t) +
    {\mathbf{k}}_t^\mathrm{T}[K(X^*,X^*)+\sigma_v^2I]^{-1}\left({\mathbf{{z}}}^*- \mathbf{m}(X^*)\right)\\
    \mathbb{V}\left[{g}_t \right]
    = k({\mathbf{x}}_t,{\mathbf{x}}_t)-{\mathbf{k}}_t^\mathrm{T}[K(X^*,X^*)+\sigma_v^2I]^{-1}{{\mathbf{k}}_t}
  \end{gather}
\end{subequations}
where ${\mathbf{k}}_t$ is the {abbreviation for} ${\mathbf{k}}(X^*,{\mathbf{x}}_t)$.
The most widely used kernel function is the \acf{SE} covariance function \cite{books/lib/RasmussenW06}, which is also used in this paper, 
\begin{align}
\label{eq:eqSEkernel}
k(\mathbf{x},\mathbf{x}') = \sigma_0^2 \exp \left[
    -\frac{1}{2} \frac{\left(\mathbf{x}-\mathbf{x}'\right)^{\rm{T}}\left(\mathbf{x}-\mathbf{x}'\right)}{l^2}\right].
\end{align}
Here, the hyperparameters $\sigma_0^2$ and  $l$ represent the prior variance of the signal amplitude and the kernel function length scale, respectively.

\subsection{\ac{GP}-based Motion Behaviors Learning --- Limited Information of Training State }

{This paper proposes a new \ac{GP}-based 
 learning method. I.e., the \acf{NSIM} behavior of a target is learned as GPs based on the Cartesian velocities.  The Cartesian positions $\xi_{t,k}$ and $\eta_{t,k}$  are converted into the Cartesian velocities as $\Delta \xi_{t-1,k} = \xi_{t,k}-\xi_{t-1,k}$ and $\Delta\eta_{t-1,k} = \eta_{t,k}-\eta_{t-1,k}$, respectively. The \ac{NSIM} models, i.e.  $\Delta\xi_{k}= {\dot{f}}^{\xi}\left( \Delta \mathbf{x}_{t-1,k} \right): \mathbb{R}^2 \rightarrow \mathbb{R}^1$ and $\Delta\eta_{k}={\dot{f}}^{\eta}\left(\Delta \mathbf{x}_{t-1,k}\right): \mathbb{R}^2\rightarrow \mathbb{R}^1$, are learned as \acp{GP}, where
$ \Delta {\mathbf{x}_{t-1,k}} \triangleq \left[\Delta \xi_{t-1,k}, \Delta \eta_{t-1,k}\right]^{\rm{T}} $. As most target maneuvers are coupled across coordinates, the learned models capture the crucial characteristics of the target behavior during maneuvers by using both Cartesian velocities as the input of the GPs.} Specifically, the motion behavior of the targets in \eqref{eq:MTT-MOTION} is rewritten as
\begin{subequations}
\begin{align}
    \Delta \xi_{t,k} &= {\dot{f}}^{\xi}\left( \Delta \mathbf{x}_{t-1,k} \right)\\
    \xi_{t+1, k} &= \xi_{t, k} + \Delta \xi_{t, k}   + {v}_{t,k}^{\xi}\\
   \Delta\eta_{t,k} &= {\dot{f}}^{\eta}\left(\Delta \mathbf{x}_{t-1,k}\right)\\
    \eta_{t+1, k} &= \eta_{t, k} + \Delta \eta_{t, k}   + {v}_{t,k}^{\eta}.
    \end{align}
\end{subequations}
The proposed GP-based method for prediction shown in Fig. \ref{fig3} involves offline training and online prediction steps. The training data is used offline to capture the statistical characteristics of the motion behavior of the targets, which are then transferred to online tracking for state prediction.
\begin{itemize}[leftmargin=1.5mm]
    \item {\textbf{Offline training}}:
    To train the Cartesian velocity motion functions, we take 
training data sets $ \mathcal{D}_{\xi}:= \{ \Delta X^*, \Delta {\Xi}^* \}$ and $ \mathcal{D}_{\eta}:= \{ \Delta X^*, \Delta {\Gamma}^*\}$. Specifically, the training inputs for both \acp{GP} are the same as 
\begin{align*}
    \Delta X^* &= [\Delta{\mathbf{x}}_{0}^*, \Delta{\mathbf{x}}_{2}^*, \dots, \Delta{\mathbf{x}}_{N-1}^*]
    \end{align*}
while the output data are different and set as 
\begin{align*}
    \Delta {\Xi}^* &= [\Delta{\xi}_{1}^*, \Delta{\xi}_{2}^*, \dots, \Delta{\xi}_{N}^*]^{T}\\
    \Delta {\Gamma}^* &= [\Delta{\eta}_{1}^*, \Delta{\eta}_{2}^*, \dots, \Delta{\eta}_{N}^*]^{T}.
\end{align*}
Here, $N$ denotes the number of training samples. The Cartesian $X$/$Y$ velocity motion  functions in the offline training process, i.e. $\dot{f} ^{*\xi}(\Delta {\mathbf{x}^{*}_{n}})$ and $ {\dot{f}}^{*\eta}(\Delta {\mathbf{x}^{*}_{n}})$,  are modeled as two \acp{GP} shown in \eqref{eq:eqDGPR1} and \eqref{eq:eqDGPR2}.
\begin{align}
\label{eq:eqDGPR1}
{\dot{f}}^{*\xi}(\Delta {\mathbf{x}^{*}_{n}}) & \sim \mathcal{GP}_{\xi} \left({m}_{\xi}(\Delta {\mathbf{x}^{*}_{n}}), k_{\xi}(\Delta {\mathbf{x}^{*}_{n}}, (\Delta {\mathbf{x}^{*}_{n}})') \right)\\
\label{eq:eqDGPR2}
{\dot{f}}^{*\eta}(\Delta {\mathbf{x}^{*}_{n}}) & \sim \mathcal{GP}_{\eta}  \left(
{m}_{\eta}(\Delta {\mathbf{x}^{*}_{n}}),
k_{\eta}(\Delta {\mathbf{x}^{*}_{n}}, (\Delta {\mathbf{x}^{*}_{n}})') \right).
\end{align}
Here, $\Delta {\mathbf{x}^{*}_{n}}$ and $\Delta {\mathbf{x}^{*}_{n}}'$ are the input variables,  the squared exponential  (SE) defined in \eqref{eq:eqSEkernel} are used to calculate $k_{\xi}(\Delta {\mathbf{x}^{*}_{n}}, (\Delta {\mathbf{x}^{*}_{n}})')$ and $k_{\eta}(\Delta {\mathbf{x}^{*}_{n}}, (\Delta {\mathbf{x}^{*}_{n}})')$ with the same  hyperparameters $\sigma_0^2$ and  $l$. 
Examples of the learned \acp{GP} are shown in Fig. \ref{fig4}, where the motion behavior of the target follows the gradual coordinated turns (GCT) model, defined in Section V. The turn rate of the target follows the left and coordinated right turns ($\pm 15^\circ$/s for 10s). In Fig. \ref{fig4}, the $X$ and $Y$ axes represent the input of the \ac{GP}, i.e., Cartesian velocities, and the Z axis represents the prediction value. The training samples are shown in blue points. The orange-edge and green surfaces represent the mean value of the predictive Gaussian distribution for the $X$-axis and $Y$-axis velocity, respectively, which are joint functions of current Cartesian velocities.
\begin{figure}[!t]
	\centering
	\includegraphics [width=0.47\textwidth]{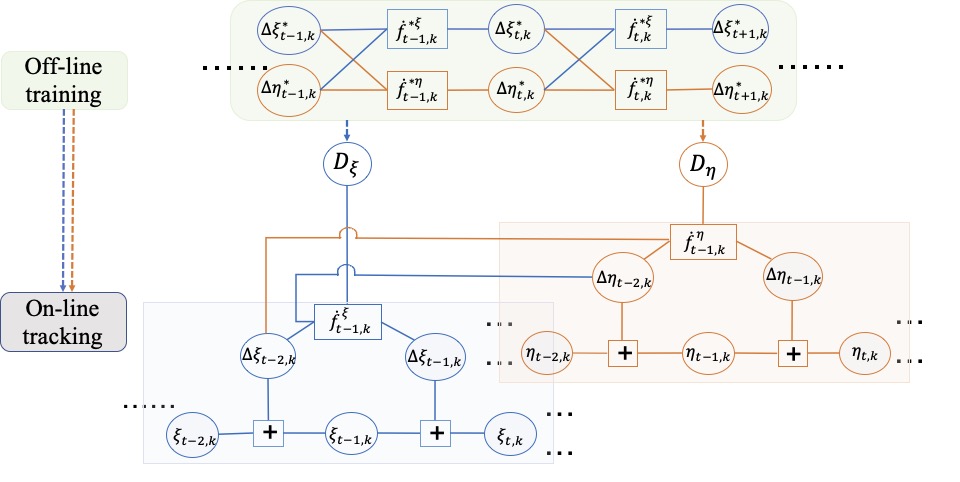}
	\caption{\footnotesize{The factor graph of the proposed 
	\ac{GP}-based learning algorithm for state prediction mainly includes offline training and online prediction processes. Legend: Circles – variable nodes; Squares – factor nodes. The process for the prediction on $X$-axis is marked in blue. The process for the prediction on $Y$-axis is marked in orange.}}
	\label{fig3}
\end{figure} 
\item  {\textbf{Online Prediction}}: 
In the following, we describe the estimation of the $X$-axis position  ${\xi}$. The estimation of the $Y$-axis position ${\eta}$ is done similarly and so is not discussed.  The estimated parameters at time $t-1$ include the Cartesian position and velocity estimates represented as $\big\{\hat{\mathbf{x}}_{t-1,k} = \left[\hat{\xi}_{t-1,k},\hat{\eta}_{t-1,k}\right]^{\rm{T}}, \widehat{\Delta \mathbf{x}}_{t-2,k}=\left[\widehat{\Delta \xi}_{t-2,k},\widehat{\Delta \eta}_{t-2,k}\right]^{\rm{T}} \big\}$.
The predictive distribution  of  the  tracking  state ${\xi}_{t,k}$ is partitioned as
\begin{equation}
\begin{aligned} 
&p ({\xi}_{t,k} \mid \hat{\mathbf{x}}_{t-1,k} \widehat{\Delta \mathbf{x}}_{t-2,k}, \mathcal{D}_{\xi}  )\\
&\qquad=\iint p ({\xi}_{t,k} \mid \hat{\xi}_{t-1,k}, {\Delta{\xi}}_{t-1,k}) \\
& \qquad\qquad \times p({\Delta{\xi}}_{t-1,k} \mid  \dot{f}_{t-2,k}^{\xi}, \mathcal{D}_{\xi}  )d{\Delta{\xi}}_{t-1,k} d \dot{f}_{t-2,k}^{\xi}.
\end{aligned}
\label{eq:GP_X_prediction}
\end{equation}
Here,  $\dot{f}_{t-2,k}^{\xi} \triangleq \dot{f}^{\xi}(\widehat{\Delta \mathbf{x}}_{t-2,k})$. The \acp{pdf} in \eqref{eq:GP_X_prediction} are further calculated in \eqref{eq:eqpdf1} -- \eqref{eq:eqpdf3},
\begin{align}
\label{eq:eqpdf1}
    \begin{split}
    &p ({\xi}_{t,k} \mid \, \hat{\xi}_{t-1,k}, {\Delta{\xi}}_{t-1,k})
    = \mathcal{N}\left({\xi}_{t,k};\,\hat{\xi}_{t-1,k}+\Delta{\xi}_{t-1,k}, Q_{\xi}\right) \end{split}\\
     %    \label{eq:eqpdf2}
     %&p ( \dot{f}_{t-2,k}^{\xi}\mid \mathcal{D}_{\xi}  ) =\mathcal{N}\left( \bar{\mu}_{{\xi}}(\widehat{\Delta \mathbf{x}}_{t-2,k}), \bar{K}_{\xi}(\widehat{\Delta \mathbf{x}}_{t-2,k})\right)\\
      \label{eq:eqpdf3}
         \begin{split}
          &p({\Delta{\xi}}_{t-1,k} \mid  \dot{f}_{t-2,k}^{\xi}, \mathcal{D}_{\xi} )\\ &\qquad\qquad=\mathcal{N}\left(\Delta{\xi}_{t-1,k};\,\bar{\mu}_{{\xi}}(\widehat{\Delta \mathbf{x}}_{t-2,k}), \bar{K}_{\xi}(\widehat{\Delta \mathbf{x}}_{t-2,k})\right)
     \end{split}
\end{align}
where the predictive mean $\bar{\mu}_{{\xi}}(\widehat{\Delta \mathbf{x}}_{t-2,k})$ and covariance $\bar{K}_{\xi}(\widehat{\Delta \mathbf{x}}_{t-2,k})$ are calculated in \eqref{eq:delta_mean} and \eqref{eq:delta_var}, respectively.
\begin{gather}
\label{eq:delta_mean}
    \bar{\mu}_{\xi}(\widehat{\Delta \mathbf{x}}_{t-2,k})=
      \bm{k}_{\xi} \left[{K}_{{\xi}}
    +\sigma_v^2 I \right]^{-1}\Delta {\Xi}^*\\
\label{eq:delta_var}
\bar{K}_{\xi}(\widehat{\Delta \mathbf{x}}_{t-2,k})={k}_{\xi}(\widehat{\Delta \mathbf{x}}_{t-2,k},\widehat{\Delta \mathbf{x}}_{t-2,k})
-\bm{k}_{\xi}^\mathrm{T}[{K}_{{\xi}}+\sigma_v^2I]^{-1}{\bm{k}_\xi}
\end{gather}
where $\bm{k}_{\xi}$ and ${K}_{{\xi}}$ are the {abbreviations for}  $\bm{k}_{\xi}(\Delta{X}^*,\widehat{\Delta \mathbf{x}}_{t-2,k})$ and ${K}_{{\xi}}(\Delta{X}^*,\Delta{X}^*)$.
\end{itemize}

\begin{figure}[!t]
		\begin{center}
		\subfigure[] {\label{fig:3Da}
			\includegraphics[width=0.4\textwidth]{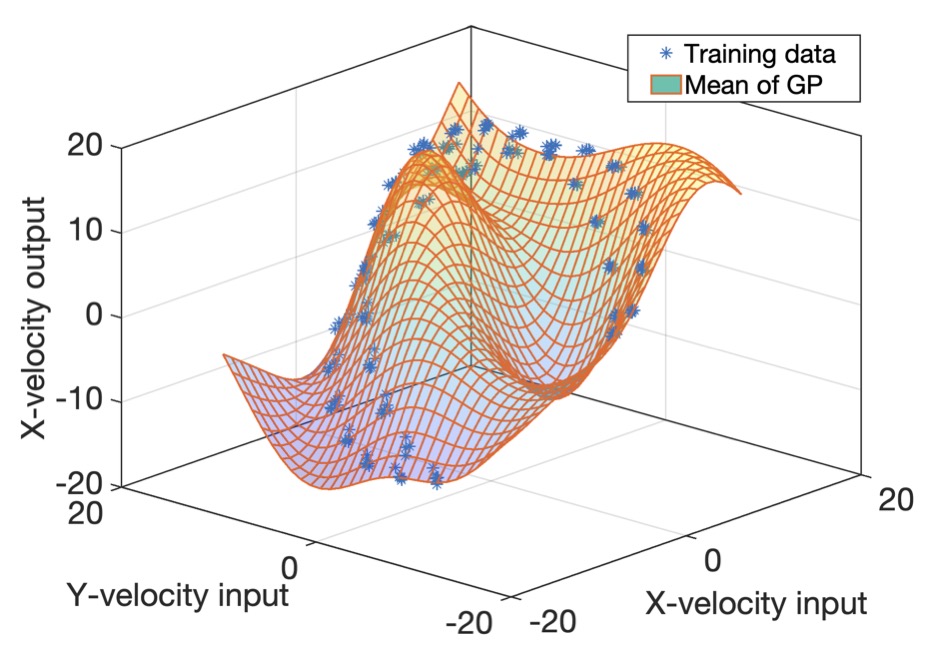}}
\subfigure[] {\label{fig:3Db}
	\includegraphics[width=0.4\textwidth]{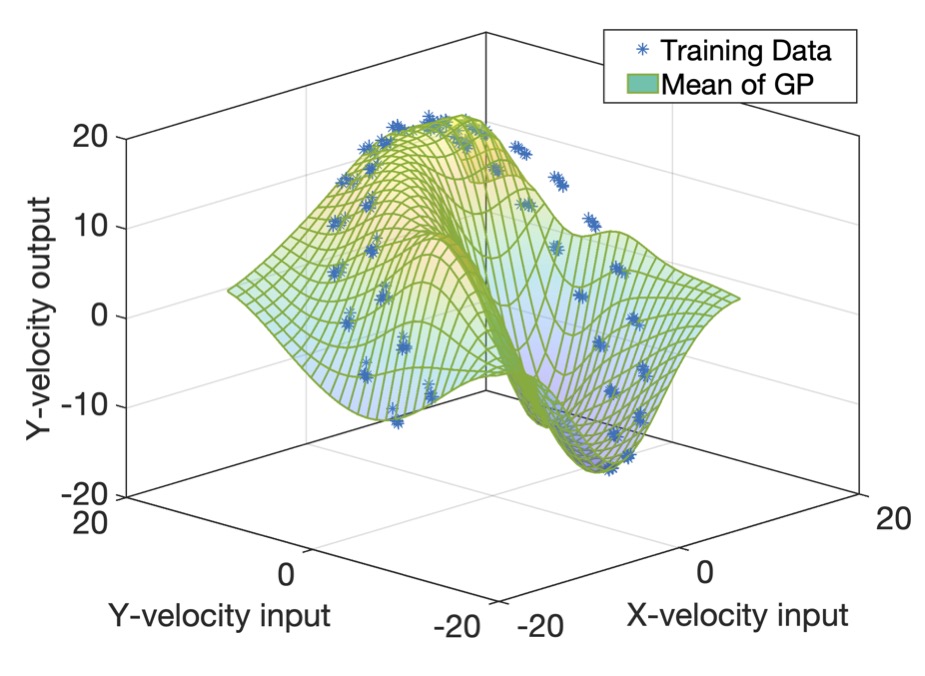}}
	\caption{\footnotesize{Example of two-dimensional \ac{GPR}.
	Legend: Points -- training samples; Surface -- the mean function of the trained \acp{GP}. (a) \ac{GP} for $X$-axis velocity. (b) \ac{GP} for $Y$-axis velocity.}}
	\label{fig4} 
	\end{center}
\end{figure}

\section{Incorporation of Learned \ac{GP}-based  \ac{NSIM} Model into Target Tracking}

For nonlinear and non-Gaussian motion and observation models, the \ac{pdf} of the hidden states in \eqref{eq:eq_marginal_pdf} and \eqref{eq:GP_X_prediction} cannot be evaluated in closed form. Therefore, particle-based implementation algorithms are commonly used for target tracking. The proposed GP-based \ac{NSIM} model learning method is plugged into the \ac{PF} and PF-based \acf{BP} algorithms in this section. When the parametric motion model is unavailable, particles are drawn based on the learned \acp{GP}. The learned GP-based PF-MP implementation for \ac{MTT} is summarized in Algorithm I. The GP-based PF for \ac{STT} is implemented the same way without executing Step 5, i.e. the loopy \ac{BP}. {Note that the data association for \ac{MTT} is obtained by running an iterative BP algorithm following \cite{6978890}. Therefore, the proposed algorithm is still valid when the target number is unknown and time varying.}

\begin{algorithm}[t]
    \caption{Learned GP based PF-\ac{BP} filter}
  \begin{algorithmic}[1]
    \REQUIRE Measurement model $h(\cdot)$ and training data $\mathcal{D}$. 
    \STATE \textbf{Off-line training}: 
    
    The X/Y-axis velocity motion  functions  are modeled as two \acp{GP}, 
    
{${\dot{f}}^{*\xi}(\Delta {\mathbf{x}^{*}_{n}})  \sim \mathcal{GP}_{\xi} \left({m}_{\xi}(\Delta {\mathbf{x}^{*}_{n}}), k_{\xi}(\Delta {\mathbf{x}^{*}_{n}}, (\Delta {\mathbf{x}^{*}_{n}})') \right)$,

${\dot{f}}^{*\eta}(\Delta {\mathbf{x}^{*}_{n}})  \sim \mathcal{GP}_{\eta}  \left(
{m}_{\eta}(\Delta {\mathbf{x}^{*}_{n}}), 
k_{\eta}(\Delta {\mathbf{x}^{*}_{n}}, (\Delta {\mathbf{x}^{*}_{n}})') \right).$}

\vspace{1mm}
    \STATE \textbf{On-line tracking}: 
    \FOR{$t = 1 : T$}
    
      \STATE Draw the particle vector

      $\mathcal{F}_{t,k}^{\{m\}} = [{\xi}_{t,k}^{\{m\}},{\eta}_{t,k}^{\{m\}}, {\Delta \xi}_{t-1,k}^{\{m\}},{\Delta \eta}_{t-1,k}^{\{m\}}]^{\rm{T}}$ as:

    ${\Delta\xi}_{t-1,k}^{\{m\}} \sim \mathcal{N}( \bar{\mu}_{f_{\xi}}(\Delta \mathbf{x}_{t-2,k}^{\{m\}}), \bar{K}_{\xi}(\Delta \mathbf{x}_{t-2,k}^{\{m\}}))$,

    ${\Delta\eta}_{t-1,k}^{\{m\}} \sim \mathcal{N}( \bar{\mu}_{f_{\xi}}(\Delta \mathbf{x}_{t-2,k}^{\{m\}}), \bar{K}_{\eta}(\Delta \mathbf{x}_{t-2,k}^{\{m\}}))$,
    
${\xi}_{t,k}^{\{m\}} \sim \mathcal{N}({\Delta \xi}_{t-1,k}^{\{m\}}+ {\xi}_{t-1,k}^{\{m\}}, Q_{\xi})$,

${\eta}_{t,k}^{\{m\}} \sim \mathcal{N}({\Delta \eta}_{t-1,k}^{\{m\}}+ {\eta}_{t-1,k}^{\{m\}}, Q_{ \eta})$.

\vspace{1mm}
Calculate the approximation of the predictive state as

$p(\mathbf{x}_{t,k}\mid {Y}_{1:t-1}) = \sum_{m=1}^{M} w_{t-1, k}^{\{m\}}\delta(\mathbf{x}_{t,k}-\mathbf{x}_{t,k}^{\{m\}})$.

\vspace{1mm}
 { \STATE Execute the loopy \ac{BP} by passing messages repeatedly,
%\vspace{0.01mm}

    $\mu_{k \rightarrow \kappa } = \frac{\psi_{k}(\kappa)}{1+\sum_{\kappa' \neq \kappa, \kappa'>0 }\psi_{k}(\kappa') \upsilon_{\kappa' \rightarrow k}}$,
    
     $\upsilon_{\kappa \rightarrow k } = \frac{1}{1+\sum_{k' \neq k, k'>0 }\psi_{k}(\kappa') \mu_{k' \rightarrow \kappa}}$.

\vspace{1mm}

Then, approximate  the  marginal association probability as
\vspace{0.05mm}
    $\hat{p}(a_{t,k} = \kappa \mid Y_t ) = \frac{\psi_{k}(\kappa) \upsilon_{\kappa \rightarrow k}}{\sum_{\kappa'} \psi_{k}(\kappa') \upsilon_{\kappa' \rightarrow k}}$.}

\vspace{1mm}
     \STATE Update associated weights,
 \vspace{0.05mm}    
 
      $\tilde{w}_{t,k}^{\{m \}} \propto \sum_{\kappa=1}^{{\mathcal{K}}}  w_{t-1,k}^{\{m \}}
    \frac{p(\mathbf{y}_{t,\kappa}\mid \mathbf{x}^{\{m\}}_{t,k})\hat{p}(a_{t,k} = \kappa \mid Y_t )p(\mathbf{x}^{\{m\}}_{t,k}\mid \mathbf{x}^{\{m\}}_{t-1,k})}{q(\mathbf{x}^{\{m\}}_{t,k}\mid \mathbf{x}^{\{m\}}_{t-1,k}, \mathbf{y}_{t,\kappa})}$.
    
    \vspace{1mm}
     \STATE Weights normalize and resampling: The particles with negligible weights are replaced by new particles in the proximity of the particles with higher weights.
       %\vspace{-4mm}
        % \begin{multicols}{2}

       %\end{multicols}  
    \ENDFOR
  \end{algorithmic}
\end{algorithm}

\subsection{Draw Particles based on Learned \acp{GP} }

The  particle vectors for the $k$-th target at the preceding time  slot $t-1$ are represented as
\begin{align*}
\begin{split}
    \mathcal{F}_{t-1,k}^{\{m\}} &= 
    \left[
       {\mathbf{x}}_{t-1,k}^{\{m\}}, {\Delta \mathbf{x}}_{t-2,k}^{\{m\}}
   \right]^{\rm{T}}
   =  \left[
      {\xi}_{t-1,k}^{\{m \}}, {\eta}_{t-1,k}^{\{m\}},
       {\Delta\xi}_{t-2,k}^{\{m\}},
       {\Delta \eta}_{t-2,k}^{\{m\}}
    \right]^{\rm{T}}
    \end{split}
\end{align*}
with the corresponding posterior weights $w_{t-1,k}^{\{m\}}$, $m = 1,\dots, M$. The particle vector $\mathcal{F}_{t-1,k}^{\{m\}}$ consists of the Cartesian position particles on the $X$-axis and $Y$-axis, i.e. $\mathbf{x}_{t-1,k}^{\{m\}} = \left[ {\xi}_{t-1,k}^{\{m \}}, {\eta}_{t-1,k}^{\{m\}}\right]^{\rm{T}}$,  and the Cartesian velocity particles on the $X$-axis and $Y$-axis, i.e. $\Delta \mathbf{x}_{t-2,k}^{\{m\}} = \left[\Delta {\xi}_{t-2,k}^{\{m \}}, \Delta {\eta}_{t-2,k}^{\{m\}}\right]^{\rm{T}}$.  The posterior probability of $k$-th target's positions and velocities at time $t-1$ is approximated as
\begin{equation}
    \begin{aligned}
     p(\mathcal{F}_{t-1,k}\mid Y_{1:t-1})
    \approx   \sum_{m=1}^M 
  \mathcal{F}_{t-1,k}^{\{m\}}{w}_{t-1,k}^{\{m\}}.  
\end{aligned}
\end{equation}
%\begin{align}
%     p(\mathcal{F}_{t-1,k}\mid Y_{1:t-1})
%    \approx  \sum_{m=1}^M 
%    \begin{bmatrix}
%       {\xi}_{t-1,k}^{\{m\}}\\{\eta}_{t-1,k}^{\{m\}}\\
%      {\Delta \xi}_{t-2,k}^{\{m\}}\\{\Delta \eta}_{t-2,k}^{\{m\}}
%    \end{bmatrix}{w}_{t-1,k}^{\{m\}}.   
%\end{align}

The new particles at time $t$ are drawn based on the importance densities represented as $\pi(\cdot)$. Specifically, %the Cartesian velocity motion behavior in the X and Y axes, i.e. the \ac{NSIM} model, are learned as two \acp{GP}, elaborated in section II. Hence, 
the particles of the Cartesian velocities are sampled from the Gaussian distributions based on the learned \acp{GP} as 
\begin{align}
\begin{split}
    {\Delta\xi}_{t-1,k}^{\{m\}} &\sim \pi({\Delta\xi}_{t-1,k}  \mid {\Delta\mathbf{x}}_{t-2,k}^{\{m\}})= p ( \dot{f}^{\xi}({\Delta \mathbf{x}}_{t-2,k}^{\{m\}})  \mid \mathcal{D})\\
    &=\mathcal{N}\left( \bar{\mu}_{{\xi}}(\Delta \mathbf{x}_{t-2,k}^{\{m\}}), \bar{K}_{\xi}(\Delta \mathbf{x}_{t-2,k}^{\{m\}})\right)
\end{split}\\
\begin{split}
    {\Delta\eta}_{t-1,k}^{\{m\}} &\sim \pi({\Delta\eta}_{t-1,k}  \mid {\Delta\mathbf{x}}_{t-2,k}^{\{m\}})= p ( \dot{f}^{\eta}({\Delta \mathbf{x}}_{t-2,k}^{\{m\}}) \mid \mathcal{D}  )\\
    &=\mathcal{N}\left( \bar{\mu}_{{\eta}}(\Delta \mathbf{x}_{t-2,k}^{\{m\}}), \bar{K}_{\eta}(\Delta \mathbf{x}_{t-2,k}^{\{m\}})\right). 
\end{split}
\end{align}
Here, $\mathcal{D}$ is the training data. The mean and covariance are calculated based on \eqref{eq:delta_mean} and  \eqref{eq:delta_var}, respectively.
Next, the position particles  ${\xi}_{t,k}^{\{m\}}$ and ${\eta}_{t,k}^{\{m\}}$ are sampled according to \eqref{eq:eqdiff1-1} and \eqref{eq:eqdiff2-2} as
\begin{align}
\label{eq:eqdiff1-1}
{\xi}_{t,k}^{\{m\}} &\sim q({\xi}_{t,k}  \mid {\Delta \xi}_{t-1,k}^{\{m\}}, {\xi}_{t-1,k}^{\{m\}}) = \mathcal{N}\left({\Delta \xi}_{t-1,k}^{\{m\}}+ {\xi}_{t-1,k}^{\{m\}}, Q_{\xi}\right)\\
\label{eq:eqdiff2-2}
{\eta}_{t,k}^{\{m\}} &\sim q({\eta}_{t,k}  \mid {\Delta \eta}_{t-1,k}^{\{m\}}, {\eta}_{t-1,k}^{\{m\}}) = \mathcal{N}\left({\Delta \eta}_{t-1,k}^{\{m\}}+ {\eta}_{t-1,k}^{\{m\}}, Q_{ \eta}\right).
\end{align}
Then, the prediction of the position state at time $t$ is approximated as,
\begin{align}
\label{eq:eqpredictionprob}
    p(\mathbf{x}_{t,k} \mid {Y}_{1:t-1}) = \sum_{m=1}^{M} w_{t-1, k}^{\{m\}}\delta(\mathbf{x}_{t,k}-\mathbf{x}_{t,k}^{\{m\}})
\end{align}
where $\mathbf{x}_{t,k}^{\{m\}} = \left[{\xi}_{t,k}^{\{m\}} ,{\eta}_{t,k}^{\{m\}}\right]^{\mathrm{T}}$. 

\subsection{Data Association}

In the \ac{PF}-based \ac{BP}, the  marginal association probabilities, $ {p}(a_{n,k} = \kappa \mid Y_t)$ and ${p}(b_{n,m} = k \mid Y_t)$, are calculated based on the loopy \ac{BP}, i.e. updating the messages between factor nodes $a_{t,k}$ and $b_{t,k}$, as shown in Fig. \ref{fig1}. Specifically, the two sets of messages are represented as $\mu_{k \rightarrow \kappa }$ and $\upsilon_{\kappa \rightarrow k }$ to denote the message passed from node $a_{t,k}$ to node $b_{t,\kappa}$ and the message passed from the node $b_{t,\kappa}$  to the node $a_{t,k}$, respectively. {The messages are updated as \cite{6978890}
\begin{align}
\label{eq:eqMP1-1}
    \mu_{k \rightarrow \kappa } = \frac{\psi_{k}(\kappa)}{1+\sum_{\kappa' \neq \kappa, \kappa'>0 }\psi_{k}(\kappa') \upsilon_{\kappa' \rightarrow k}}\\
    \label{eq:eqMP2-2}
        \upsilon_{\kappa \rightarrow k } = \frac{1}{1+\sum_{k' \neq k, k'>0 }\psi_{k}(\kappa') \mu_{k' \rightarrow \kappa}}
\end{align}
where the factor $\psi_{k}(\kappa)$ indicates the association probability calculated based on the \ac{LR} as
\begin{equation}
\label{association_probability1}	
\psi_k(\kappa)=	
\begin{cases}
\frac{P_D \int p\left(\mathbf{y}_{t,\kappa}\mid \, { {\mathbf{x}}_{t,k}}\right)  p(\mathbf{x}_{t,k}\mid {Y}_{1:t-1}) d{\mathbf{x}_{t,k}}}{\lambda_{\text{fa}}\left[1- 
P_D\int p(\mathbf{x}_{t,k}\mid {Y}_{1:t-1})d{\mathbf{x}_{t,k}}\right]},&\kappa = 1,..., \mathcal{K}_t,\\
1, &\kappa=0
\end{cases}
\end{equation}
%\begin{equation}
%\label{association_probability1}	
%\psi_k(\kappa)=	
%\begin{cases}
%\frac{P_D \int p\left(\mathbf{y}_{t,\kappa}\mid \, {\mathbf{x}_{t,k}}\right)p(\mathbf{x}_{t,k}| \,{\mathbf{x}}_{t-1,k}) d{\mathbf{x}_{t,k}}}{\lambda_{\text{fa}}\left[1-P_D  +\int p(\mathbf{x}_{t,k}|{\mathbf{x}}_{t-1,k}) d{\mathbf{x}_{t,k}}\right]}, &\kappa = 1,..., \mathcal{K}_t,\\
%1, &\kappa=0,
%\end{cases}
%\end{equation}
where $\kappa=0$ when no measurement is correctly associated to $k$-th target, and the predictive distribution $ p(\mathbf{x}_{t,k}\mid {Y}_{1:t-1})$ is approximated as  \eqref{eq:eqpredictionprob}. 
The messages are passed between the latent association variables until convergence, i.e. \eqref{eq:eqMP1-1} and \eqref{eq:eqMP2-2} are repeatedly processed until the conditions of  convergence is satisfied \cite{DBLP:journals/jmlr/IhlerFW05}. Finally, the approximations of the  marginal association probabilities  are obtained as}
\begin{align}
\label{eq:eqa_cala}
    \hat{p}(a_{t,k} = \kappa \mid Y_t ) = \frac{\psi_{k}(\kappa) \upsilon_{\kappa \rightarrow k}}{\sum_{\kappa'} \psi_{k}(\kappa') \upsilon_{\kappa' \rightarrow k}}\\
    \label{eq:eqa_calb}
     \hat{p}(b_{t,k} = k \mid Y_t ) = \frac{ \mu_{k \rightarrow  \kappa}}{\sum_{\kappa'} \psi_{k}(\kappa') \upsilon_{\kappa' \rightarrow k}}.
\end{align}

\subsection{PF-based State Estimation }
After these marginal association probabilities in \eqref{eq:eqa_cala} and \eqref{eq:eqa_calb} are obtained, the marginal  posterior \ac{pdf} $p(\mathbf{x}_{t,k}\mid Y_{1:t})$ in \eqref{eq:eq_marginal_pdf} is then approximated as
\begin{align}
\label{eq:eqpredictionprob111}
    p(\mathbf{x}_{t,k} \mid {Y}_{1:t}) = \sum_{m=1}^{M} w_{t, k}^{\{m\}}\delta(\mathbf{x}_{t,k}-\mathbf{x}_{t,k}^{\{m\}})
\end{align}
where the updated weights $w_{t, k}^{\{m\}}$ are calculated by
\begin{gather}
\begin{split}
    \tilde{w}_{t,k}^{\{m \}}\propto w_{t-1,k}^{\{m \}} \sum_{\kappa=1}^{\mathcal{K}_t}  
    p(\mathbf{y}_{t,\kappa}\mid \mathbf{x}^{\{m\}}_{t,k})\hat{p}(a_{t,k} = \kappa \mid Y_t )%p(\mathbf{x}^{\{m\}}_{t,k}\mid {Y}_{1:t-1})
    \end{split}\\
 {w}_{t,k}^{\{m \}} = \frac{ \tilde{w}_{t,k}^{\{m \}} }  { \sum_{m=1}^M\tilde{w}_{t,k}^{\{m \}} }, \;\; m = 1,\dots,M.
\end{gather}

%the posterior estimation of targets' positions and   velocities are approximated by multiplying particles with updated weightsbased on \eqref{eq:PF_MP}. T

Based on the key idea of the \ac{PF} \cite{978374, 1232326}, the \ac{MAP} estimation  of hidden states for the $k$-th target at time $t$, i.e. $\mathcal{F}_{t,k}$, are approximated as
\begin{align}
\begin{split}
     \hat{\mathcal{F}}_{t,k} &=  \left[
       \hat{\mathbf{x}}_{t,k} , \widehat{\Delta \mathbf{x}}_{t-1,k}
    \right]^{\rm{T}}  =  \left[
       \hat{\xi}_{t,k},\hat{\eta}_{t,k},
       \widehat{\Delta \xi}_{t-1,k},\widehat{\Delta \eta}_{t-1,k}
    \right]^{\rm{T}}\\ 
    & = \arg \max \left[p(\mathcal{F}_{t,k}\mid Y_{1:t})\right]\\  &
     \approx \arg \max \bigg\{ \sum_{m=1}^M 
    \left[
       {\xi}_{t,k}^{\{m\}},{\eta}_{t,k}^{\{m\}},
      {\Delta \xi}_{t-1,k}^{\{m\}},{\Delta \eta}_{t-1,k}^{\{m\}}
    \right]^{\rm{T}} {w}_{t,k}^{\{m\}}\bigg\}.  
\end{split}
\label{eq:PF_MP}
\end{align}
%\begin{align}
%\begin{split}
%     &\hat{\mathcal{F}}_{t,k}=  \begin{bmatrix}
%       \hat{\mathbf{x}}_{t,k} \\ \widehat{\Delta \mathbf{x}}_{t-1,k}\\
%    \end{bmatrix}   =  \begin{bmatrix}
%       \hat{\xi}_{t,k}\\\hat{\eta}_{t,k}\\
%       \widehat{\Delta \xi}_{t-1,k}\\\widehat{\Delta \eta}_{t-1,k}
%    \end{bmatrix} = \arg \max \left[p(\mathcal{F}_{t,k}\mid Y_{1:t})\right],\\
%    & \approx \arg \max \bigg\{ \sum_{m=1}^M 
%    \begin{bmatrix}
%       {\xi}_{t,k}^{\{m\}}\\{\eta}_{t,k}^{\{m\}}\\
%      {\Delta \xi}_{t-1,k}^{\{m\}}\\{\Delta \eta}_{t-1,k}^{\{m\}}
 %   \end{bmatrix}{w}_{t,k}^{\{m\}}\bigg\}.  
%\end{split}
%\label{eq:PF_MP}
%\end{align}
%Here, $w_{t,k}^{\{m\}}$ is the weight corresponding to the $m$-th particle vector, and $m = 1, \dots, M$. 
where $\hat{\mathbf{x}}_{t,k} = \left[\hat{\xi}_{t,k}, \hat{\eta}_{t,k} \right]^{\rm{T}}$  represents the estimates of the $k$-th target's position, and $ \widehat{\Delta \mathbf{x}}_{t-1,k} = \left[\widehat{\Delta \xi}_{t-1,k}, \widehat{ \Delta \eta}_{t-1,k} \right]^{\rm{T}}$ represents the estimates of $k$-th target's Cartesian velocities between $t-1$ and $t$. 

{Based on the above descriptions, the implementation of the proposed learned GP based PF-\ac{BP} filter for target tracking is summarized in Algorithm I.}

\section{Simulation Results}
This section introduces the \acf{DSSM} used to generate simulation data and different tracking maneuvering scenarios in the first two subsections.
Then, the performance comparisons for \ac{STT} and \ac{MTT} are presented.
Specifically, the proposed learned \ac{GP} method is applied for \ac{STT} over different maneuvering scenarios. The tracking performance demonstrates the validity and robustness of the proposed training method to track the targets within different surveillance regions and motion behaviors. Then, the data-driven \ac{MTT} algorithm is compared with the oracle PF with fully known system information and the \ac{IMM-PF} methods that know the correct DA \cite{1561886}. The numbers of the used particle are set to be the same for all compared filters.

\subsection{Dynamic State-space Model (DSSM)}

{This paper generates the simulated target trajectories and measurements following the standard curvilinear-motion kinematics model \cite{1261132}.} The state of $k$-th target at time $t$ in \eqref{eq:MTT-MOTION} and 
\eqref{eq:MTT-MEA}  is represented as ${\mathbf{x}}_{t,k} = \left[ \xi_{t,k},\eta_{t,k}, v_{t,k}, \phi_{t,k}, {\alpha}_{t,k}^t, \alpha_{t,k}^n \right]$, where  $(\xi_{t,k},\eta_{t,k})$ are the target Cartesian position, $v_{t,k}$ and $\phi_{t,k}$ represent the ground speed and the velocity heading angle, respectively. The along-track and cross-track accelerations in the horizontal plane are denoted as $\alpha_{t,k}^t$ and $\alpha_{t,k}^n$, respectively. By setting different values of $\alpha_{t,k}^t$ and $\alpha_{t,k}^n$,  the standard curvilinear-motion reduces to the following special cases \cite{1261132}:
\begin{itemize} 
    \item $\alpha_{t,k}^t=0$, $\alpha_{t,k}^n = 0$ \textemdash \, \text{rectilinear}, \acf{CV} motion,
     \item $\alpha_{t,k}^t \neq 0$, $\alpha_{t,k}^n = 0$ \textemdash \text{ rectilinear, accelerated motion} (\ac{CA} motion if $\alpha_{t,k}^t$ is a  constant value),
     \item $\alpha_{t,k}^t =  0$, $\alpha_{t,k}^n \neq 0$ \textemdash \text{ circular, constant-speed motion} (\acf{CT} motion if $\alpha_{t,k}^n$ is a  constant value).
\end{itemize}

%A horizontal 2-D standard curvilinear motion kinematics model is considered for simplicity and clarity, but our method is equally applicable to more complex state space models. The geometry of targets’ motion with clutter is shown in Fig. \ref{fig2}.
%\begin{figure} [!t]
%	\centering
%	\includegraphics [width=7cm,height=5.5cm]{image/Fig.1 Geometry of 2D target motion.pdf}
%	\caption{\footnotesize{Geometry of 2D target motion.}}
%	\label{fig2}
%\end{figure}

%\subsubsection{Measurement Model}

%{Both} linear and 
The non-linear measurement model considered in this paper  is defined as
\begin{equation}
    %\mathbf{H_2}: \quad 
    h(\mathbf{x}_{t,k}) = \begin{bmatrix}
    r_{t,k}\\ \phi_{t,k}
    \end{bmatrix}
    = \begin{bmatrix}
   \sqrt{\xi_{t,k}^2 + \eta_{t,k}^2}\\ \tan^{-1}(\xi_{t,k}, \eta_{t,k}) 
    \end{bmatrix}  + \mathbf{e}_{t,k}
    \label{eq:eqH2}
\end{equation}
 where the  inverse tangent is the four-quadrant inverse tangent function. {The proposed algorithm is not limited to this measurement model but is chosen for demonstration purposes.}  

\subsection{Different Maneuvering Tracking Scenarios}
The proposed training method based on the \ac{GPR} is evaluated on the following two scenarios with different motion behaviors.

\begin{itemize}
    \item $\mathbf{S_1}$ - \Ac{GCT}: 
   The turn rate of the target follows the regular left and right coordinated turns {($\alpha_{t}^t =  0$, $\alpha_{t}^n = \pm 15^\circ$/s for 10s), as shown in Fig. \ref{fig:S1a} and Fig. \ref{fig:S1b}.} The purpose of this application is to demonstrate the
performance when tracking highly maneuvering targets with regular coordinated turns. 
   % \item $\mathbf{S_2}$ - Random coordinated turns: 
   % The turn rate of the target follows the left and right coordinated turns randomly, as shown in Fig. \ref{fig:S2}. This application aims to demonstrate the performance and robustness when tracking highly maneuvering targets with random coordinated turns.
    \item $\mathbf{S_2}$ - 
    Standard curvilinear-motion model: The target motion is modeled using the standard curvilinear-motion model, as shown in Fig. \ref{fig:S3a1}. The along-track and cross-track accelerations vary randomly among three levels, i.e. low acceleration, medium acceleration, and high acceleration, which values are set as $\{0.1, 1, 10\} \mathrm{m} / \mathrm{s}^2$. The accelerations change randomly at each time slot. This application aims to demonstrate the performance when tracking maneuvering targets with more general motion behaviors.
\end{itemize}

%The number of \ac{MC} runs is represented as $N_{\text{MC}}$.

\subsection{Single-target Tracking Performance}

The \ac{RMSE} is used to evaluate the \ac{STT}  performance for $\mathbf{S_1}$ and  $\mathbf{S_2}$ scenarios. %The test time length is set to be $T_{\textbf{S1}}$ and $T_{\textbf{S2}}$ for  $\mathbf{S_1}$ and  $\mathbf{S_2}$ scenarios, respectively.
%\begin{align}
%    \mathrm{RMSE} = 
 %   \sqrt{\frac{\sum_{i=1}^{N_{\text{MC}}} \sum_{t=1}^T(\xi_{t,i}-\hat{\xi}_{t, i})^2+(\eta_{t,i}-\hat{\eta}_{t, i})^2}{T N_{\text{MC}}}}
%\end{align}
%where $T$ represents the tracking time length and is set to be $T_{\textbf{S1}}$,  $T_{\textbf{S2}}$ and $T_{\textbf{S3}}$ for the $\mathbf{S_1}$, $\mathbf{S_2}$ and $\mathbf{S_3}$ scenarios, respectively. The number of \ac{MC} runs is represented as $N_{\text{MC}}$.

\subsubsection{ Motion  scenario $\mathbf{S_1}$}

\begin{figure}[!t]
\begin{center}
\includegraphics[width=0.35\textwidth]{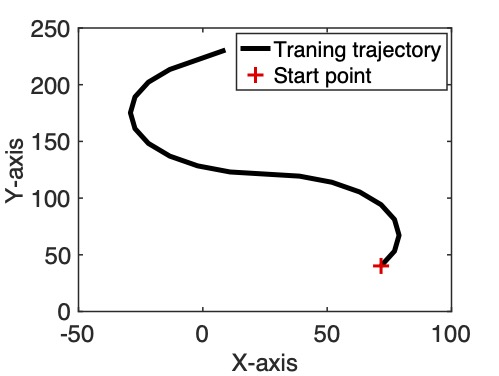}
\end{center}
\caption{Ground truth trajectory of the training data for $\mathbf{S}_1$.}
\label{fig:S1a}
\end{figure}

\begin{figure}[!htb]
	\begin{center}
	\subfigure[] {\label{fig:S1b}
			\includegraphics[width=7cm, height=5cm]{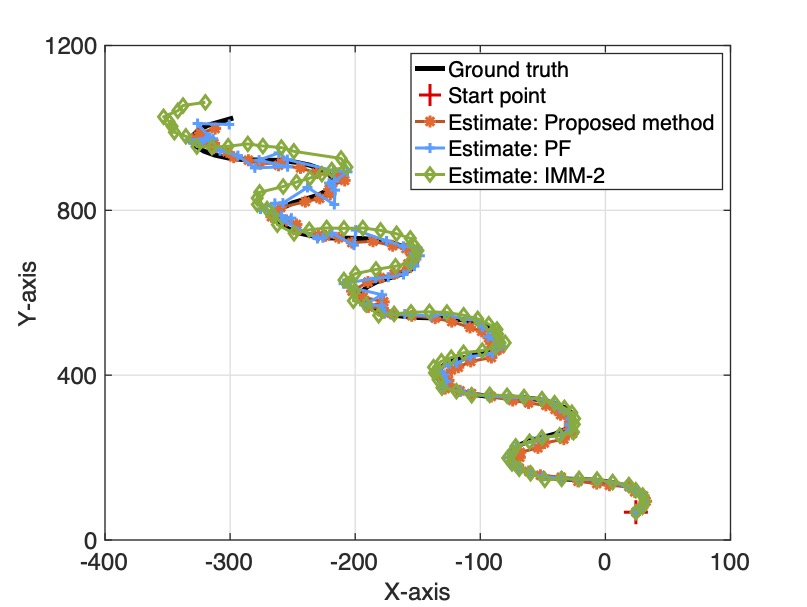}}
			\vspace{-1mm}
	\subfigure[] {\label{fig:S1b_1}
			\includegraphics[width=4cm, height=2.7cm]{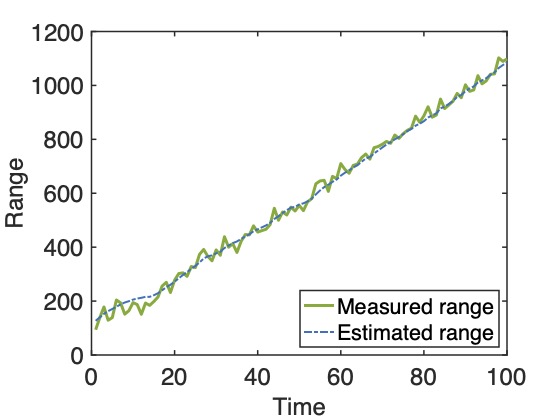}}
			\vspace{-1mm}
	\subfigure[] {\label{fig:S1b_2}
			\includegraphics[width=4cm, height=2.7cm]{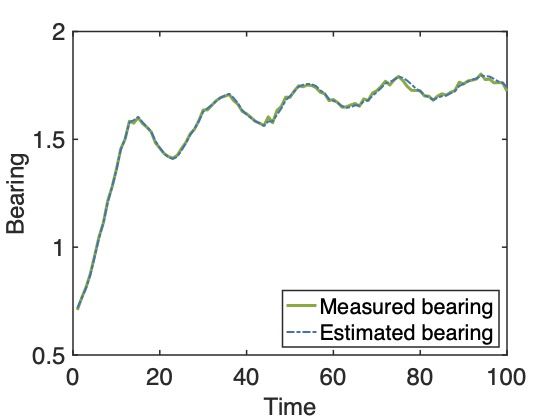}}
			\vspace{-1mm}
	\subfigure[] {\label{fig:S1c}
			\includegraphics[width=5cm, height=2.7cm]{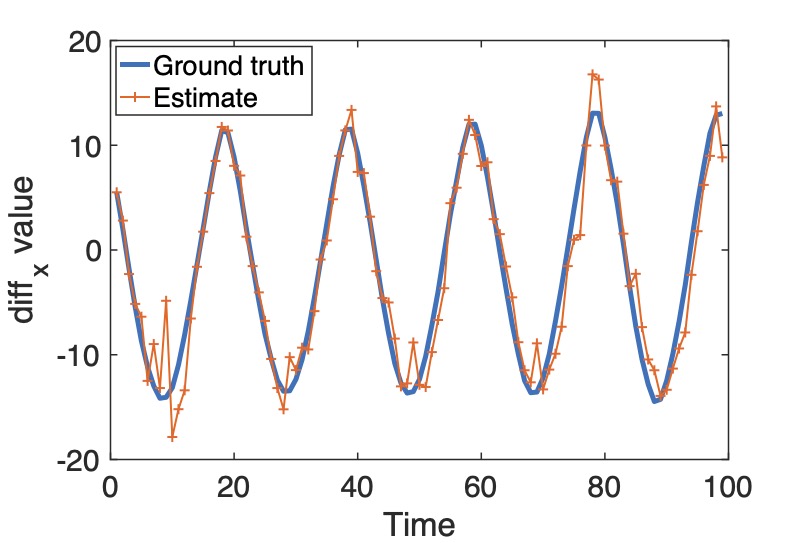}}
			\vspace{-1mm}
		\subfigure[] {\label{fig:S1d}
			\includegraphics[width=5cm, height=2.7cm]{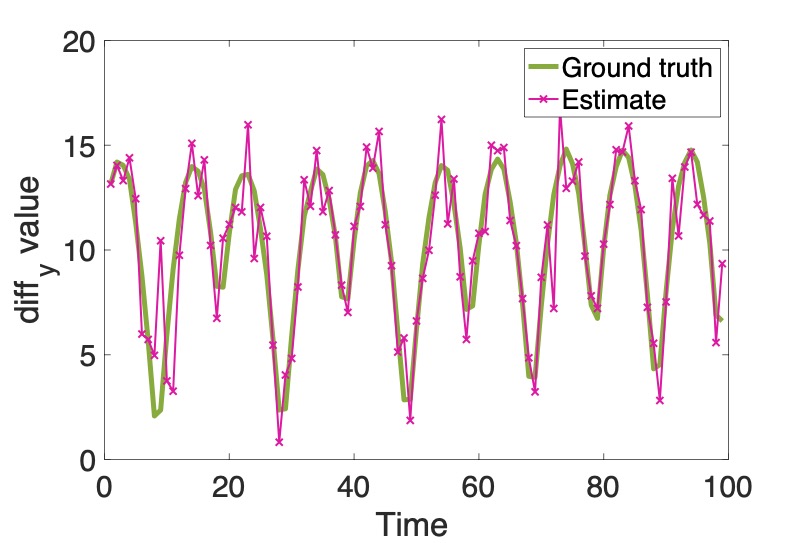}}
\caption{ Tracking performance of $\mathcal{GP}_{\textbf{S1}}$ for scenario $\mathbf{S}_1$. (a) Tracking trajectories. {(b) Measured and estimated range with time.  (c) Measured and estimated bearing with time.} (d) Estimations of X-axis Cartesian velocities with time. (e) Estimations of Y-axis Cartesian velocities with time.}
\label{fig:S1}
	\end{center}
\end{figure}

The training data length for $\mathbf{S_1}$ is $N_{\textbf{S1}} = 20$, and the trajectory is shown in Fig. \ref{fig:S1a}. The learned GP models are represented as $\mathcal{GP}_{\textbf{S1}}$.
The test length is set as $T_{\textbf{S1}} = 100$,
and the trajectory is shown in Fig. \ref{fig:S1b}. The number of  particles  $M_{\textbf{S1}} = 200$. 
The proposed algorithm is compared with the other two methods. The oracle PF gives the benchmark performance with full prior knowledge of the real-time coordinated turns, and the IMM-PF method represented as IMM-2 includes two coordinated turns models, i.e. constant left $15^\circ$/s turn model and constant right $15^\circ$/s turn model.

One example of the tracking performance for $\mathbf{S_1}$ is shown in Fig. \ref{fig:S1b}, the measured and estimated range and bearing are shown in  Fig. \ref{fig:S1b_1} and Fig. \ref{fig:S1b_2}, respectively.
The ground truth and estimates of Cartesian velocities obtained by the proposed algorithm, i.e. $\{ \Delta \xi_{1:T}, \widehat{\Delta \xi}_{1:T} \}$ and   $\{ {\Delta \eta}_{1:T}, \widehat{\Delta \eta}_{1:T} \}$, are shown in  Fig. \ref{fig:S1c} and Fig. \ref{fig:S1d}, respectively.  The \acp{RMSE} is obtained for 50 \ac{MC} realizations are shown in Table \ref{tab:STT-RMSE-S1}. From Fig. \ref{fig:S1} and Table \ref{tab:STT-RMSE-S1}, we arrive at the following conclusions. First, the proposed tracking method performs closely to the benchmark filter. Second, the proposed NSIM model based on Cartesian velocities are translationally invariant and the proposed algorithm can be applied to track targets within different surveillance regions $\left[ -300,100 \right]\times\left[ 0,1200 \right]$ from the surveillance region of the training data $\left[ -50,100 \right]\times\left[ 0,250 \right]$.

\begin{table}
\centering
\caption{Comparisons of  RMSE performance for $\mathbf{S_1}$}
\begin{tabular}{c|c|c|c}%{p{0.1\textwidth}|p{0.08\textwidth}|p{0.08\textwidth}|p{0.08\textwidth}}  
\hline\hline
Method & Proposed method  & Oracle  PF & $\mathrm{IMM-2}$\\ 
\hline \hline
RMSE -- $\mathbf{S_{1}}$ &   15.5414   &  $\mathbf{15.1072}$  &   33.6308 \\ 
 \hline 
\end{tabular}
\label{tab:STT-RMSE-S1}
\end{table}

\begin{figure}[!htb]
	\begin{center}
	\subfigure[] {\label{fig:GCT_5}
			\includegraphics[width=4cm, height=3cm]{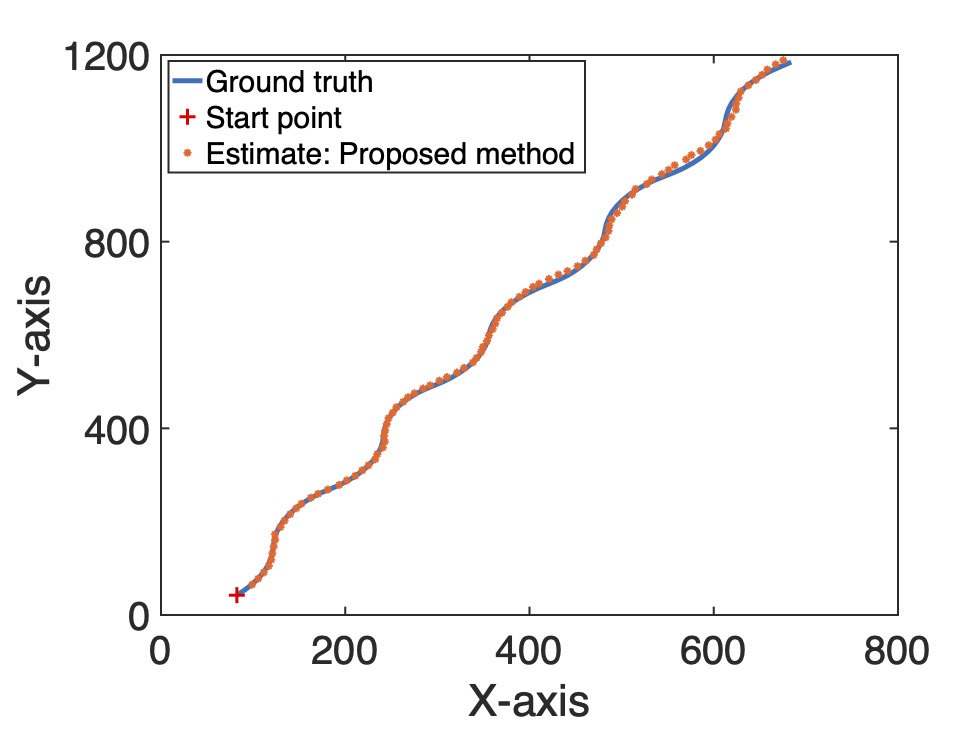}}
\subfigure[] {\label{fig:GCT_10}
			\includegraphics[width=4cm, height=3cm]{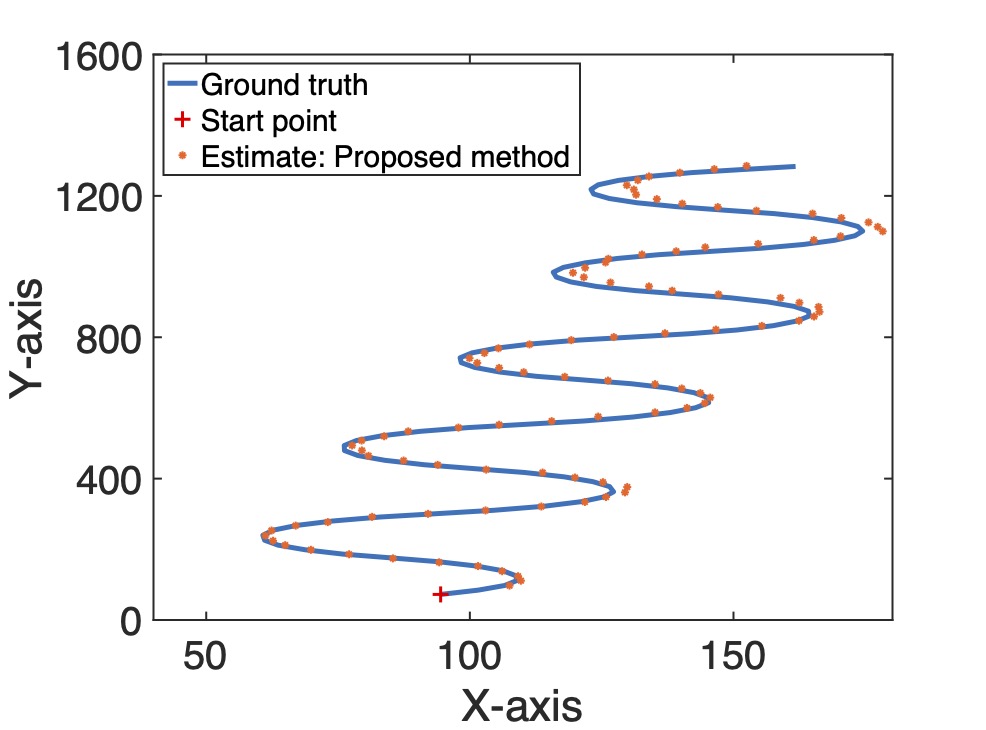}}
\subfigure[] {\label{fig:GCT_20}
			\includegraphics[width=4cm, height=3cm]{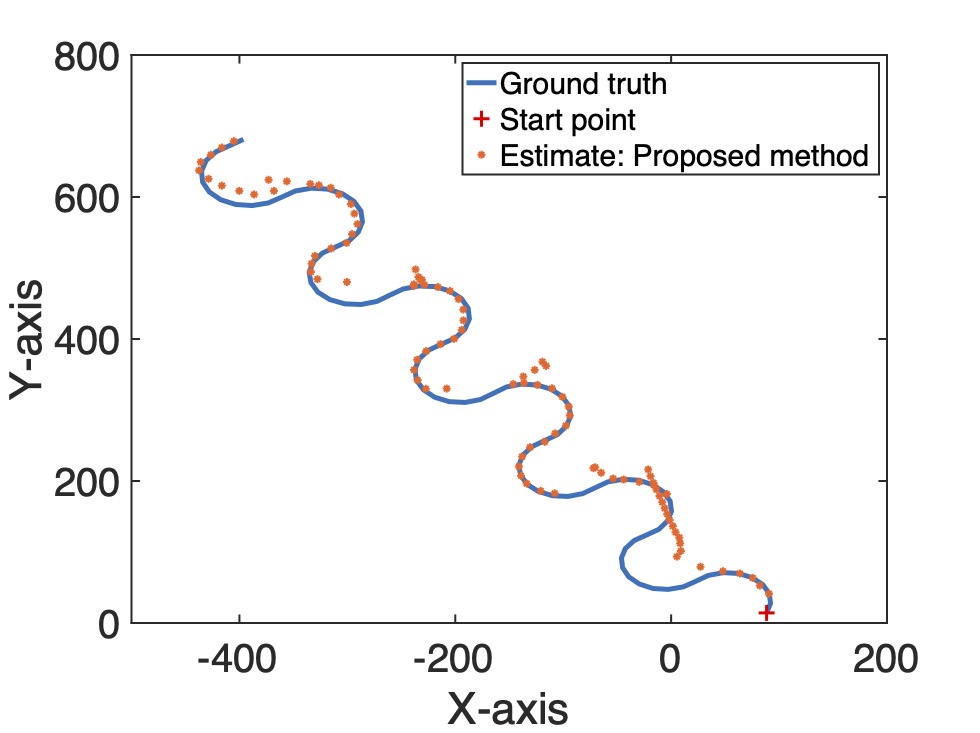}}
\subfigure[] {\label{fig:GCT_30}
			\includegraphics[width=4cm, height=3cm]{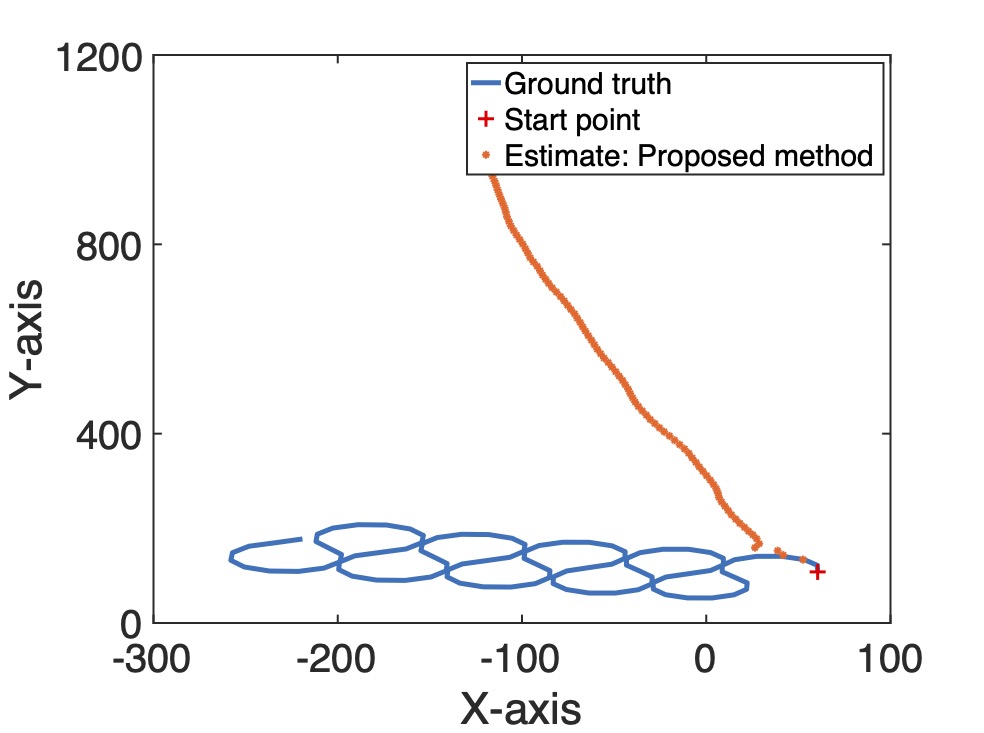}}
\caption{ Tracking performance of $\mathcal{GP}_{\textbf{S1}}$  for  \acp{GCT}  with different turn rate sets. $\mathcal{GP}_{\textbf{S1}}$ is trained based on a \ac{GCT}  data set with $\pm 15\circ$. Legend: start point of the moving trajectory.
(a) $\pm 5^\circ$. 
(b) $\pm 10^\circ$.
(c) $\pm 20\circ$. (d) $\pm 30^\circ$.}
\label{fig:GCT_different_turn}
\end{center}
\end{figure}

To evaluate the robustness of the trained motion model, the trained \acp{GP} based on  $\pm 15^\circ$ turn rate, i.e. $\mathcal{GP}_{\textbf{S1}}$, are then applied to track the targets which move following the gradual coordinated turns model but with different turn rates, i.e. $\left[ \pm 5^\circ, \pm10^\circ, \pm20^\circ, \pm30^\circ \right]$. The tracking performance is shown in Fig. \ref{fig:GCT_different_turn}. We can see that the performance is still favorable when the turn rates are set to be $\pm 5^\circ$ and $\pm 10^\circ$, i.e. those absolute values are smaller than the turn rate used for training. However, the tracking performance of the proposed method is drastically reduced when the turn rate is  $\pm 20^\circ$ and loses the tracking when the turn rate is $\pm 30^\circ$ which is very far away from the training data turn rate.

\subsubsection{Motion  scenario $\mathbf{S_2}$ }
 \begin{figure}[!t]
\begin{center}
 	\includegraphics[width=8cm, height=8cm]{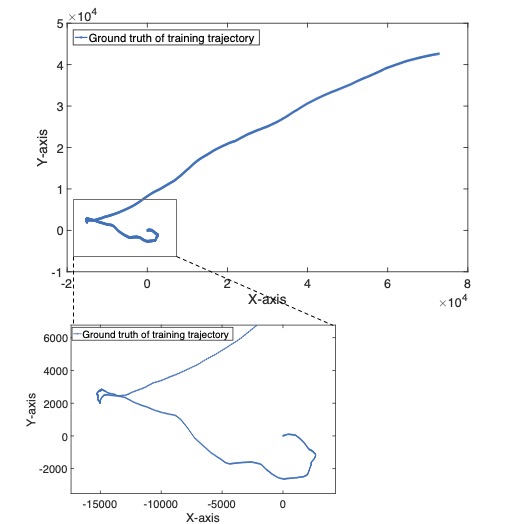}
 	\end{center}
 	\caption{{Ground truth trajectory of the training data for $\mathbf{S}_3$.}}
 	\label{fig:S3a1}
 	\end{figure}
\begin{figure}[!h]
	\begin{center}
		\subfigure[] {\label{fig:S3_a}
			\includegraphics[width=6cm, height=4.2cm]{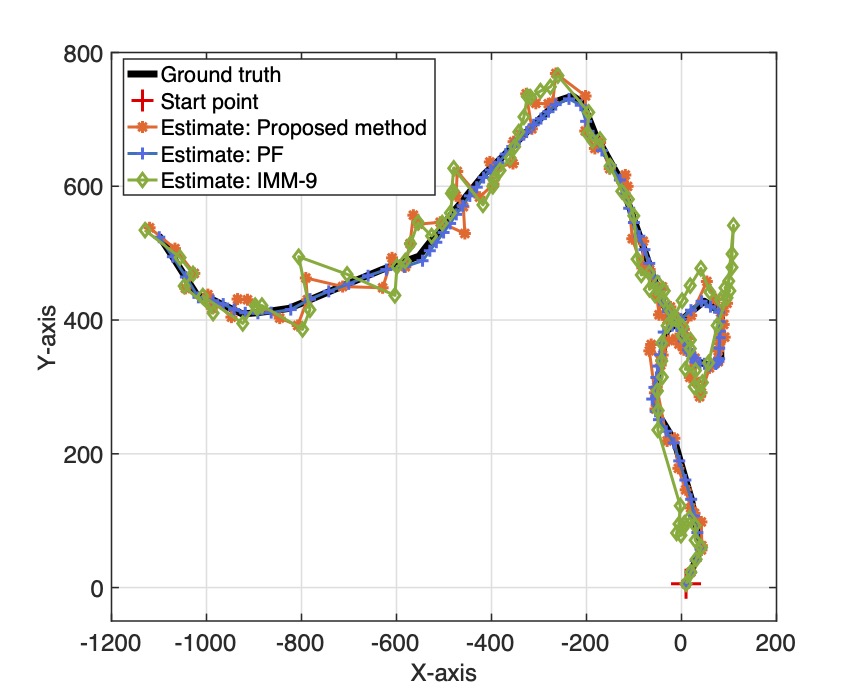}}
				\subfigure[] {\label{fig:S3c}
			\includegraphics[width=6cm, height=4.2cm]{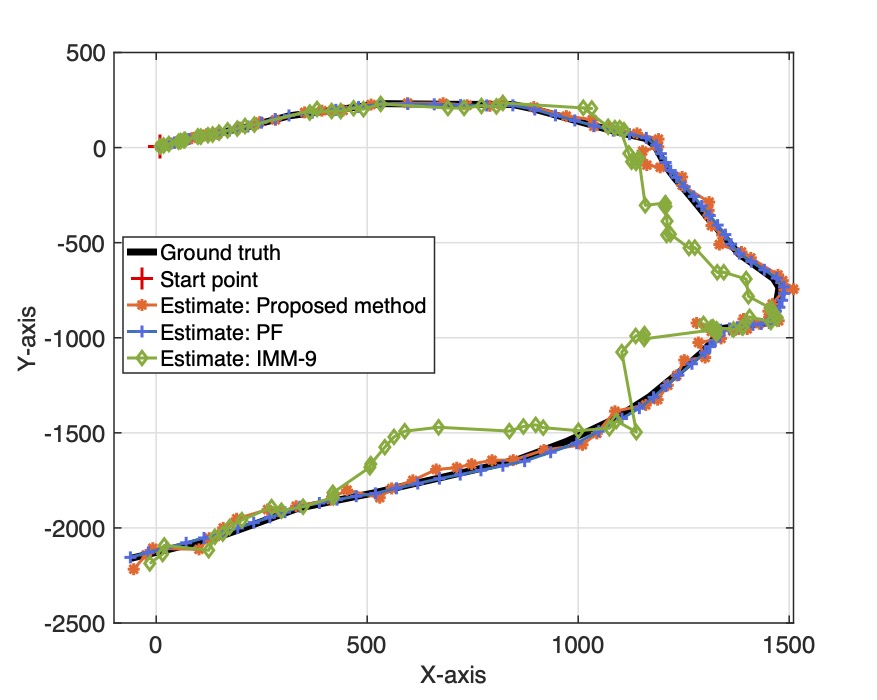}}
		\caption{{\label{fig:S3_tracking}  Tracking performance of $\mathcal{GP}_{\textbf{S3}}$. (a) Trajectory--1.  (b) Trajectory--2.  }}
	\end{center}
\end{figure}

 The training data length for $\mathbf{S_2}$ is $N_{\textbf{S2}} = 1000$, and the ground truth trajectory for training is shown in Fig. \ref{fig:S3a1}. The learned \ac{GP} models are represented as $\mathcal{GP}_{\textbf{S2}}$. The test data length and the number of particles are set as $T_{\textbf{S2}} = 100$,   $M_{\textbf{S2}} = 200$, respectively. {The ground truth trajectories for training and testing are different, i.e. not from the same dataset, but follow the same motion model.} The oracle \ac{PF} gives the benchmark performance with full prior knowledge of the real-time along-track and cross-track accelerations. {The \ac{IMM-PF} compared is represented as IMM-9. All the potential motion parameters of the test data are set as the candidates of the IMM-PF algorithm, i.e. nine motion models are set as the candidates with  ${\alpha}_{t,k}^t \in \{0.1, 1, 10\}$ and ${\alpha}_{t,k}^n\in \{0.1,  1, 10\}$.} Fig.  \ref{fig:S3_a}  and Fig.  \ref{fig:S3c} show the tracking performance comparison of two typical examples. {Note that the trajectory in Fig. \ref{fig:S3a1} used for training differs from those used for the actual test experiment, but both are generated following the same motion model.} The \ac{RMSE}
performance comparisons obtained for 50 \ac{MC} realizations are shown in Table \ref{tab:STT-RMSE-S3}. 
 From the simulation results, we can arrive at the following conclusions. Firstly, the proposed GP-based learning method has good suitability and scalability for different moving trajectory schemes. Secondly, compared with the IMM-PF, the tracking performance of the proposed method is better. Compared with the oracle PF, the performance of the proposed algorithm is similar despite the unknown motion parameters. 
\begin{table}[!htbp]
 \centering
\caption{ Comparisons of  RMSE performance for $\mathbf{S_2}$}
\begin{tabular}{c|c|c|c} %\begin{tabular}{p{0.1\textwidth}|p{0.08\textwidth}|p{0.08\textwidth}|p{0.08\textwidth}}  % repeats {c|} 18 times
\hline\hline
Method & {Proposed method}  & Oracle  PF & IMM-9\\ 
\hline \hline
RMSE -- ${\mathbf{S_{2}}}$ &   {28.9758}   &  \textbf{14.2775}   &  
{248.9746} \\ 
 \hline 
%IMM-PF &   58.8862 & 147.0132  \\
\end{tabular}
\label{tab:STT-RMSE-S3}
\end{table}

\subsection{Robustness of Learned GP models}

Given that the standard curvilinear-motion kinematics mode is a general motion model that various (noiseless) kinematic models can be derived from, in this experiment, the learned GP model $\mathcal{GP}_{\textbf{S2}}$ and IMM-9 are applied to scenarios with different motion behaviours, %e.g. $\mathbf{S}_1$, $\mathbf{S}_{2-1}$ and $\mathbf{S}_{2-2}$, 
e.g. $\mathbf{S}_1$ and $\mathbf{S}_{3}$,
to evaluate the robustness. 
$\mathbf{S}_{3}$ is defined as the target moves with random turn rate $\pm 15^\circ$s for 1s.

\begin{figure}[!htb]
	\begin{center}
				\subfigure[] {\label{fig:GP3_S1_est}
			\includegraphics[width=6.5cm, height=4.6cm]{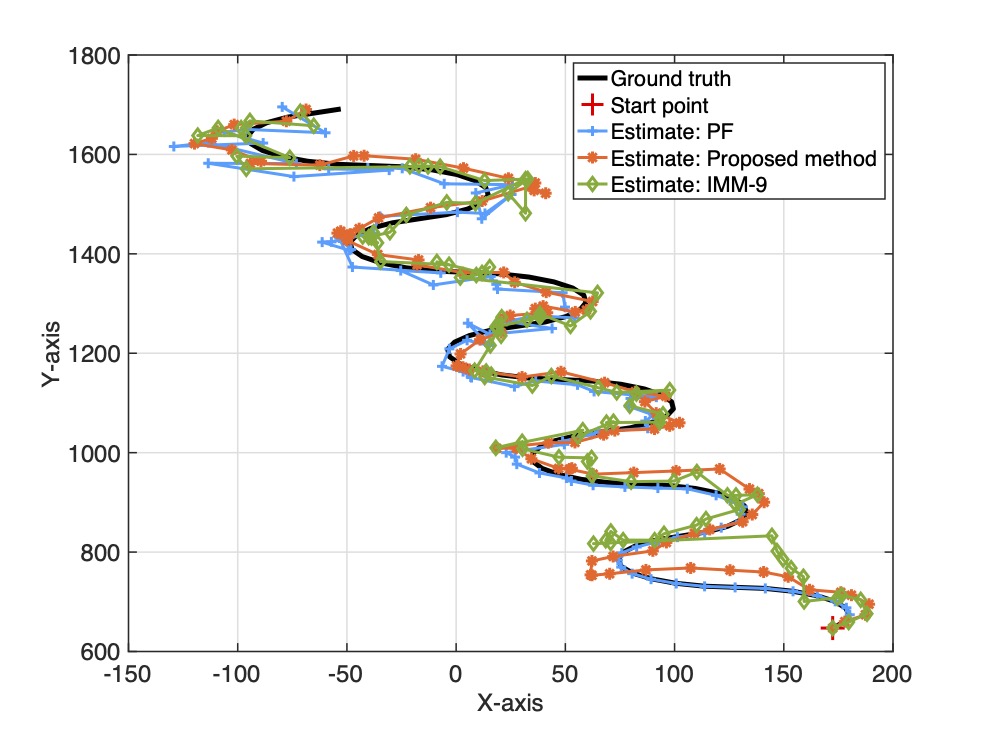}}
			%\subfigure[] {\label{fig:GP3_S2_1_est}
			%\includegraphics[width=6.5cm, height=4.6cm]{image/Estimation_trajectory_GP3_S2_1.pdf}}
\subfigure[]{\label{fig:GP3_S2_2_est} 
	\includegraphics[width=6.5cm, height=4.6cm]{
	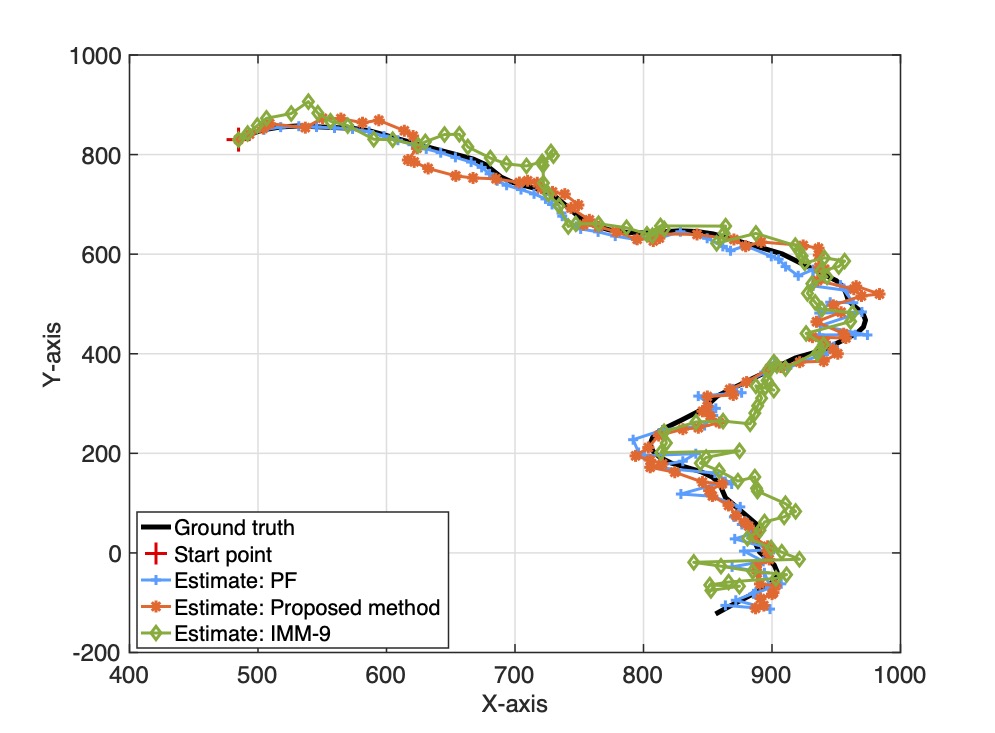}}
\caption{\label{fig:GP3} 
Tracking performance of $\mathcal{GP}_{\textbf{S2}}$. Legend: start point of the moving trajectory.
(a) ${\mathbf{S_{1}}}$. 
%(b) ${\mathbf{S_{2-1}}}$. 
(b) ${\mathbf{S_{3}}}$.}
	\end{center}
\end{figure}

One realization for different scenarios and the RMSE obtained for 50 \ac{MC} realizations are shown in Fig. \ref{fig:GP3}  and Table \ref{tab:STT-RMSE-G3}, respectively. The tracking performance comparisons demonstrate that the learned \ac{GP} models from more general training data, i.e. the learned GP model $\mathcal{GP}_{\textbf{S2}}$ has good robustness for different moving trajectories following different motion models with the same or lower maneuvering.

\begin{table}[!htbp]
 \centering
\caption{Comparisons of  RMSE performance of $\mathcal{GP}_{\mathbf{S2}}$}
\begin{tabular}{c|c|c|c}
%\begin{tabular}{p{0.11\textwidth}|p{0.08\textwidth}|p{0.08\textwidth}|p{0.08\textwidth}}
\hline\hline
Method & Proposed method  & Oracle PF & $\mathrm{IMM-9}$ \\ 
\hline \hline
RMSE -- $\mathbf{S_{1}}$  &   {15.4871}   &  $\mathbf{10.1769}$ &   17.8063 \\ 
\hline
%RMSE -- ${\mathbf{S_{2-1}}}$  &   {27.8317}   &  \textbf{8.6770}  &   {39.0177} \\ 
% \hline 
 RMSE -- ${\mathbf{S_{3}}}$  &   {18.0542}   &  $\mathbf{17.4488}$  &   {52.1975} \\ 
 \hline 
%IMM-PF &   58.8862 & 147.0132  \\
\end{tabular}
\label{tab:STT-RMSE-G3}
\end{table}

%standard curvilinear-motion kinematics model

\subsection{Multi-target Tracking Performance}

\begin{figure*}[!thb]
	\begin{center}
				\subfigure[] {\label{fig:MS2b}
			\includegraphics[width=8cm]{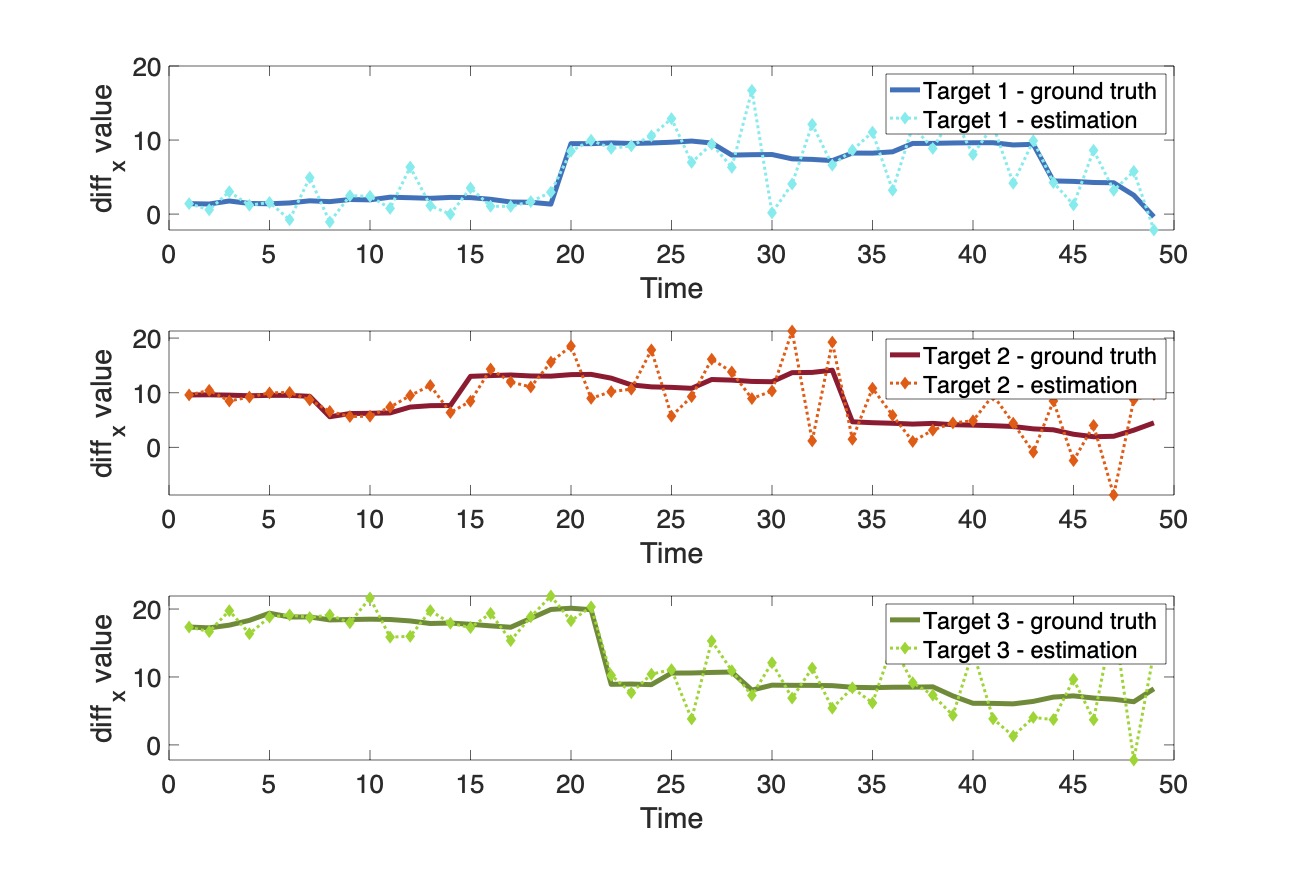}}
				\subfigure[] {\label{fig:MS2c}
			\includegraphics[width=8cm]{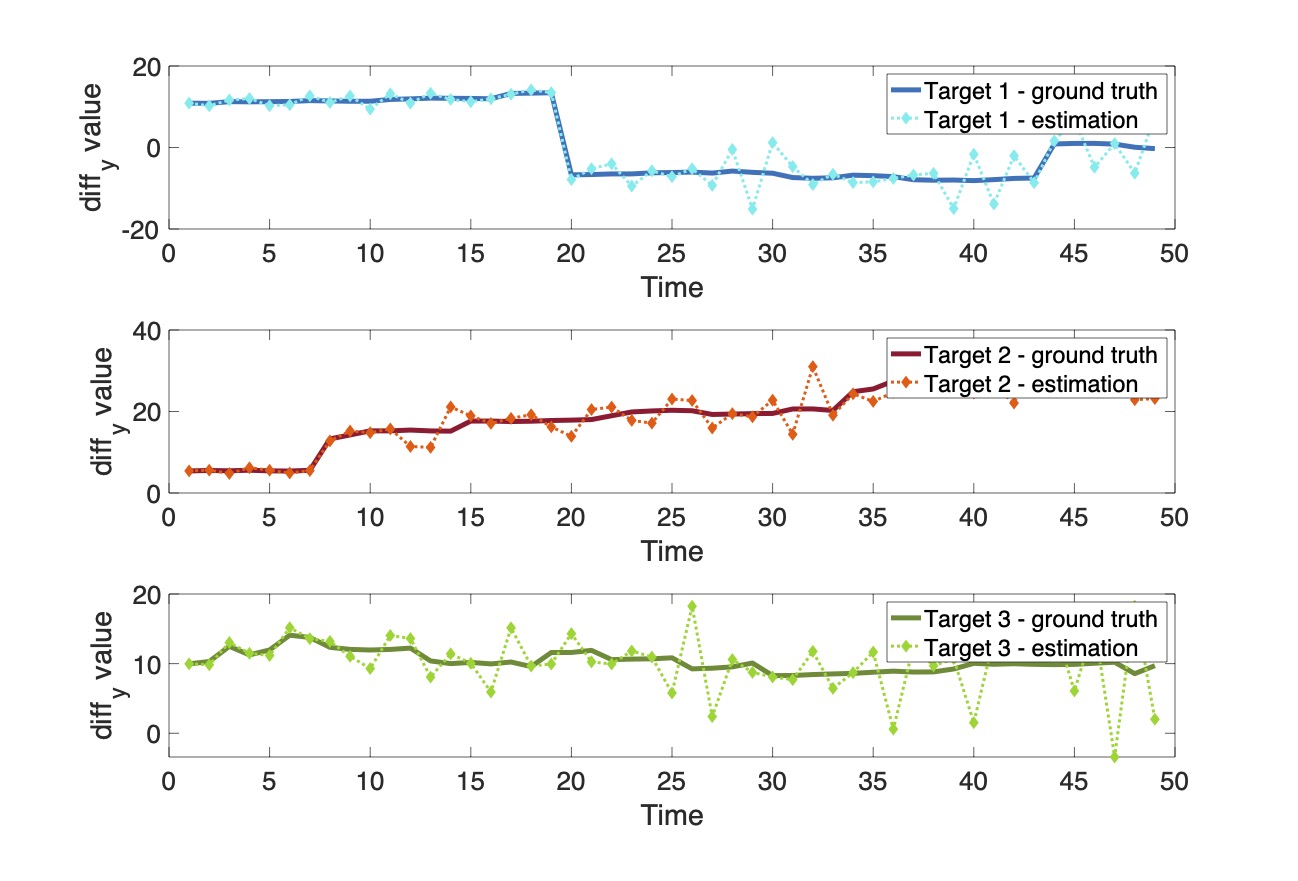}}
					\subfigure[] {\label{fig:MS2a}
			\includegraphics[width=8cm]{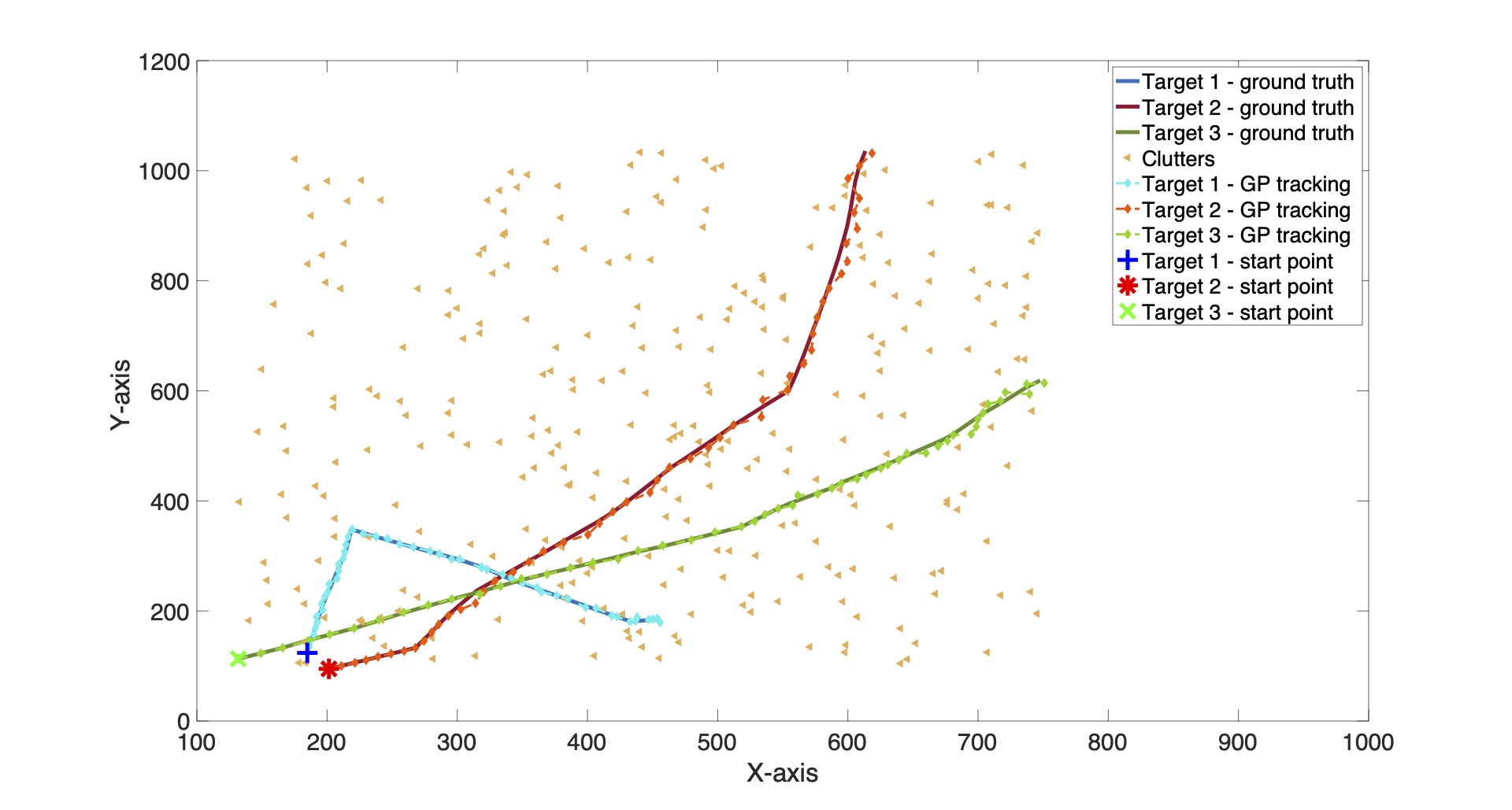}}
					\subfigure[] {\label{fig:MS2d}
			\includegraphics[width=8cm]{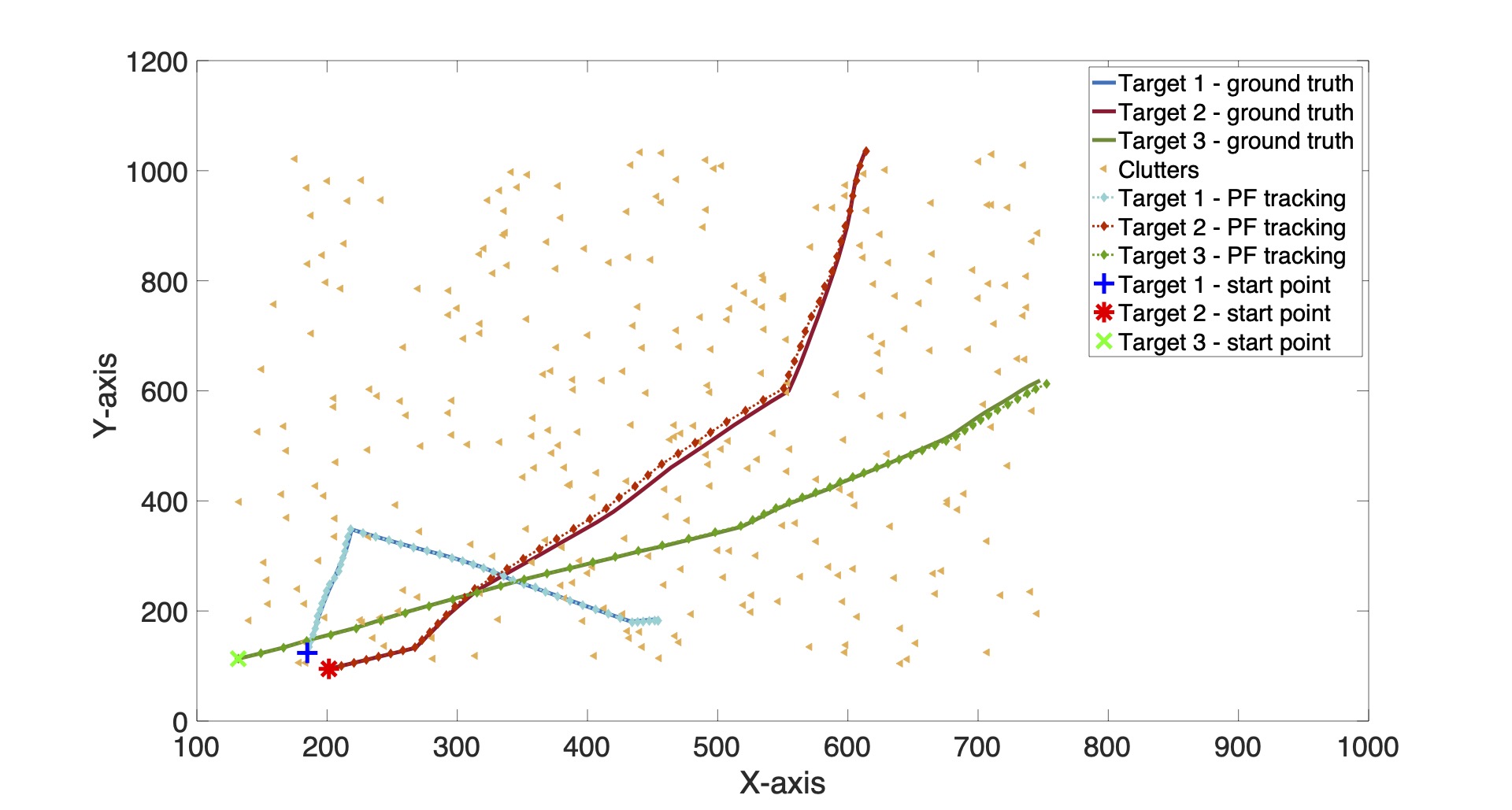}}
					\subfigure[] {\label{fig:MS2e}
			\includegraphics[width=8cm]{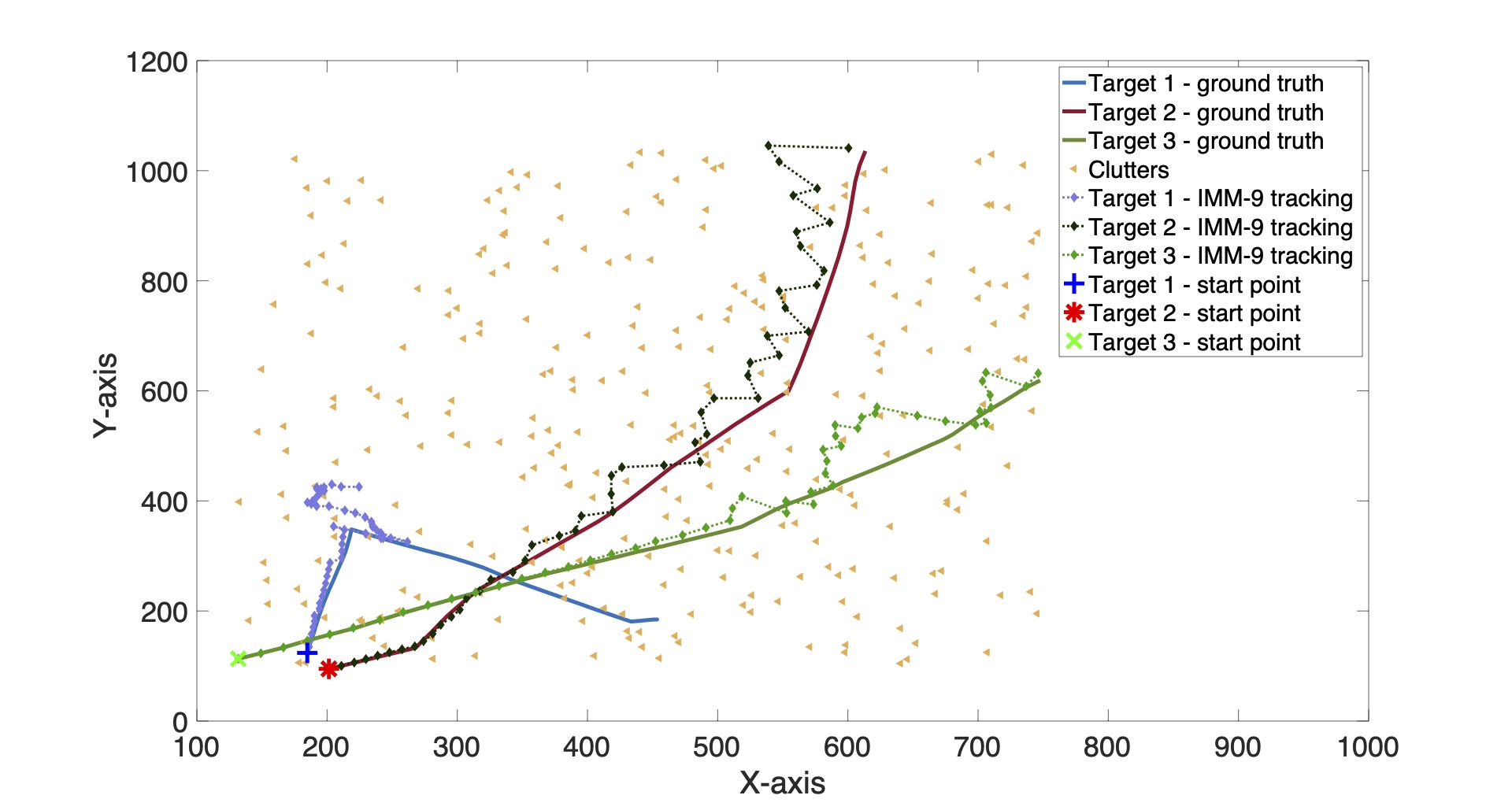}}
			\subfigure[] {\label{fig:MTTGPSAP}
			\includegraphics[width=8cm]{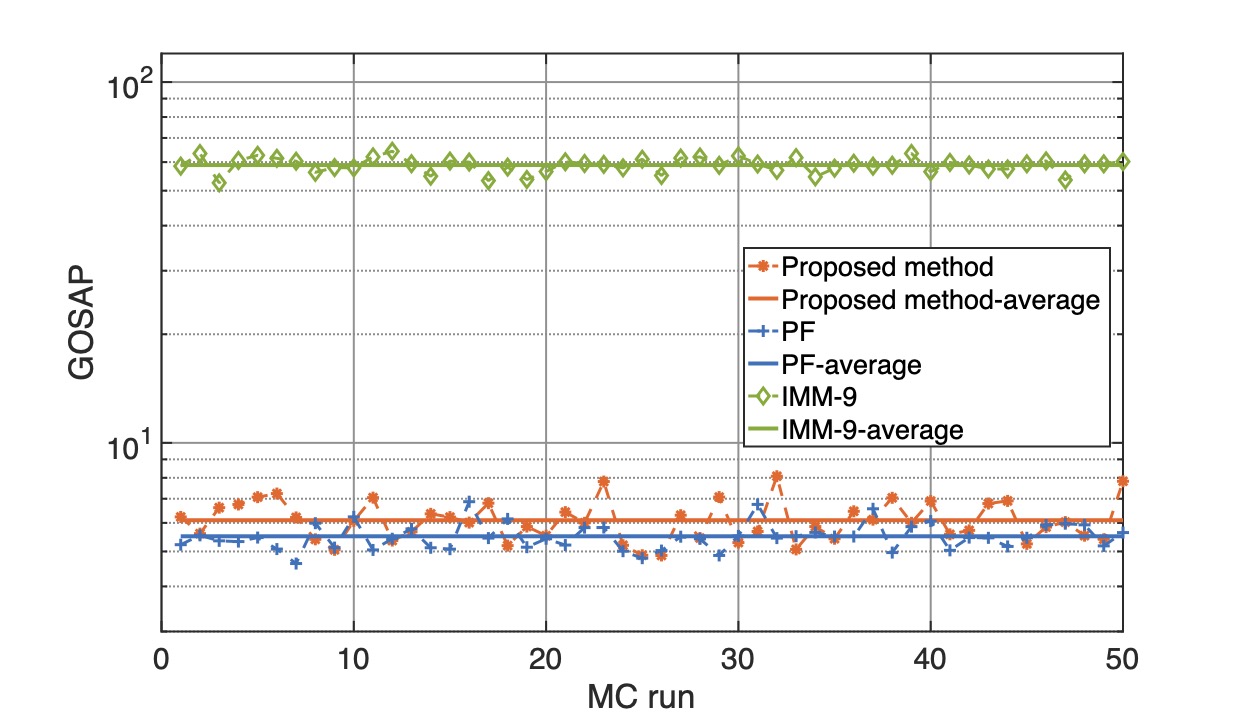}}
			%		\subfigure[] {\label{fig:MS2f}
			%\includegraphics[width=8cm]{image/Tracking_trajectory_IMM4.pdf}}
		\caption{\label{fig:MTT} Tracking performance of MTT \textbf{S2}.  
		(a) Estimations of differential X-axis position with time.
	(b) Estimations of differential Y-axis position with time.  (c) Tracking performance of the proposed method. (d) Tracking performance of the PF with real-time accelerations method. (e) Tracking performance of the IMM-9. (f) GOSAP performance of different methods for MTT. The number of targets is 3, and the motion models for all targets follow \textbf{S2}.  }
	\end{center}
\end{figure*}
%\begin{figure}[!htb]
%	\begin{center}
%	\includegraphics[width=9cm]{image/GOSAP-MTT.pdf}
%		\caption{\label{fig:MTTGPSAP} GOSAP performance of different methods for MTT. The number of targets is 3, the  motion models for all targets follow \textbf{S2}.  }
%			\end{center}
%\end{figure}

The motion behavior of the multi targets is set to follow $\mathbf{S_2}$, and the tracking time length is $T=50$. The learned \ac{GP} model $\mathcal{GP}_{\mathbf{S2}}$ is used for online tracking prediction. The number of particles is $M =500$. 
The tracking performance for one realization of four different filters is shown in Fig. \ref{fig:MS2b} to Fig. \ref{fig:MS2e}. Note that for the compared methods, i.e. the oracle \ac{PF} and the IMM-9 that includes the accurate motion model, the data association is assumed to be prior knowledge. The \ac{GOSAP} is used for evaluating the statistical tracking performance as it captures three critical factors: localization errors for correctly detected targets, number of missed detections, and number of false detections due to clutter. The \ac{GOSAP} is calculated as \cite{8009645}
\begin{equation}
d_p^{(c,\alpha)}(X,\hat{X})=\left(\min_{\pi \in \prod_{\left|\hat{X} \right| }}
\sum_{i=1}^{\left| X\right| }d^c \left( x_i,\hat{x}_{\pi(i)}\right)^{p} + \frac{c^p}{\alpha}\left(\left|\hat{X}\right| -\left|X \right| \right) \right) ^{\frac{1}{p}}.
\label{eq:gosap}
\end{equation}
Here, $X$ and $\hat{X}$ represent the states of true targets and estimations respectively, whose numbers are $\left|X \right|$ and $\left|\hat{X} \right|$. $c>0$, $0<\alpha\leq 2$ and $1\leq p <\infty$. $d(x,\hat{x})$ denote the metric for any $x,\hat{x} \in \mathbb{R}^N $ and $d^{(c)}(x,\hat{x})=\min(d(x,\hat{x}),c)$ is its cut-off metric \cite{8009645}. 
From the simulation results shown in  Fig. \ref{fig:MTTGPSAP}, we can conclude that compared  to  the  \ac{IMM}-PF methods  with  the  same number of particles, the proposed particle GP-\ac{BP} method  provides 90\%  performance improvement in the position estimation, and it  can achieve  very similar tracking performance to  the oracle PF.

 \section{conclusion}

This study proposes a new method for target tracking under unknown motion models. The method proposed is based on \ac{GPR}, which can learn target movements' \ac{NSIM} behaviors. The experiments showed that this method had satisfying accuracy and outstanding robustness for random target trajectories {and is still valid when the test data's surveillance region is outside the training data's surveillance region.} In addition, we demonstrated the immediate potential for using the proposed method as a plug-and-play prediction module in various Bayesian filtering applications.

\bibliographystyle{IEEEtran}
\bibliography{sample.bib}
\end{document}